\documentclass{aa}          
\usepackage{graphicx}
\usepackage{txfonts}
\usepackage{longtable}
\begin{document}

   \title{Asteroids seen by JWST-MIRI\thanks{This work is based on observations made with the NASA/ESA/CSA James Webb Space Telescope (JWST). The data were obtained from the ESA JWST Science Archive at \url{https://jwst.esac.esa.int/archive/}.}: Radiometric Size, Distance and Orbit Constraints}
   \author{
	   T.\ G.\ M\"{u}ller \inst{\ref{inst1}} \and                
           M.\ Micheli \inst{\ref{inst2}} \and                       
	   T.\ Santana-Ros \inst{\ref{inst3},\ref{inst4}} \and       
	   P.\ Bartczak \inst{\ref{inst5},\ref{inst6}} \and                      
           D.\ Oszkiewicz \inst{\ref{inst5}} \and                    
           S.\ Kruk \inst{\ref{inst1}}                               
          }
   \institute{   {Max-Planck-Institut f\"{u}r extraterrestrische Physik,
                 Giessenbachstra{\ss}e, Postfach 1312, 85741 Garching, Germany
                 \email{tmueller@mpe.mpg.de}\label{inst1}
                 }
                 \and
                 {ESA NEO Coordination Centre,
		 Largo Galileo Galilei, 1, 00044 Frascati (RM), Italy
                 \label{inst2}
                 }
                 \and
		 {Instituto de F{\'i}sica Aplicada a las Ciencias y las Tecnolog{\'i}as, Universidad de Alicante, San Vicente del Raspeig, 03080, Alicante, Spain
                 \label{inst3}
		 }
		 \and
		 {Institut de Ci{\`e}ncies del Cosmos (ICCUB), Universitat de Barcelona (IEEC-UB),
		 Carrer de Mart{\'i} i Franqu{\`e}s, 1, 08028 Barcelona, Spain
		 \label{inst4}
                 }
                 \and
                 {Astronomical Observatory Institute, Faculty of Physics,
                 A. Mickiewicz University, S\l oneczna 36, 60-286 Pozna\'n, Poland
                 \label{inst5}
                 }
		 \and
		 {Instituto Universitario de F{\'i}sica Aplicada a las Ciencias y las Tecnolog{\'i}as (IUFACyT).
                 Universidad de Alicante, Ctra. San Vicente del Raspeig s/n. 03690 San Vicente del Raspeig, Alicante, Spain
                 \label{inst6}
                 }
          }
   \date{Received ; accepted }
\abstract{
Infrared measurements of asteroids are crucial for the determination
of physical and thermal properties of individual objects, and for the
understanding of the small-body populations in the solar system as a
whole. But standard radiometric methods can only be applied if the orbit
of an object is known, hence its position at the time of the observation.
With JWST-MIRI observations the situation will change and many unknown,
often very small, solar system objects will be detected. Later orbit
determinations are difficult due to the faintness of the objects and
the lack of dedicated follow-up concepts. We present MIRI observations
of the outer-belt asteroid (10920) 1998~BC1 and an unknown object,
detected in all 9 MIRI bands in close apparent proximity to (10920).
We developed a new method "STM-ORBIT" to interpret the multi-band measurements
without knowing the object's true location. The power of the new technique
is that it determines the most-likely helio- and observer-centric distance and
phase angle ranges, allowing to do a radiometric size estimate.
The application to the MIRI fluxes of (10920)
was used to validate the method. It leads to a confirmation of known radiometric
size-albedo solution and puts constraints on the asteroid's location and orbit in agreement with
its true orbit. To back up the validation of the method, we obtained additional groundbased
lightcurve observations of (10920), combined with Gaia data, which indicate a
very elongated object (a/b $\ge$ 1.5), with a spin-pole
at ($\lambda$, $\beta$)$_{ecl}$ = (178$^{\circ}$, +81$^{\circ}$), with an
estimated error of about 20$^{\circ}$, and a rotation period of
4.861191 $\pm$ 0.000015\,h. A thermophysical study of all available
JWST-MIRI and WISE measurements leads to a size of 14.5 - 16.5\,km (diameter of
an equal-volume sphere), a geometric albedo p$_{V}$ between 0.05 and 0.10,
and a thermal inertia in the range 9 to 35 (best value 15)\,J\,m$^{-2}$s$^{-0.5}$K$^{-1}$.
For the newly discovered MIRI object, the STM-ORBIT method revealed a size of
100-230\,m. The new asteroid must be on a low-inclination orbit (0.7$^{\circ}$ $<$ i $<$ 2.0$^{\circ}$)
and it was located in the inner main-belt region during JWST observations.
A beaming parameter $\eta$ larger than 1.0 would push the size even below
100 meter, a main-belt regime which escaped IR detections so far. These kind
of MIRI observations can therefore contribute to formation and evolution 
studies via classical size-frequency studies which are currently limited to
objects larger than about one kilometer in size. 
We estimate that MIRI frames with pointings close to the ecliptic
and only short integration times of a few seconds will always include
a few asteroids, most of them will be unknown objects.}

\authorrunning{M\"uller et al.}
\titlerunning{Radiometric Size, Distance and Orbit Determination}
\maketitle
%

\section{Introduction}
\label{sec:intro}

The radiometric method is widely used to determine physical and thermal properties of
small atmosphereless objects in the Solar System \citep[e.g.][]{Delbo2015}. Measurements in
the thermal infrared (IR) are combined with reflected light properties (usually represented
by an object's H, G1, G2 values\footnote{H is the object's absolute magnitude, G1 and G2 describe
the shape of the phase function \citep{Muinonen2010}.})
to find size-albedo solutions which explain the 
visual magnitudes and the IR fluxes simultaneously. If sufficient good-quality IR measurements
are available, it is also possible to determine surface properties, like the roughness,
the thermal inertia or thermal conductivity. A detailed modelling of the temperature distribution
on the surface even allows to put constraints on spin or shape properties \citep[e.g.][]{Mueller2017}.
Different thermal models are available, like the Standard Thermal Model (STM; \cite{Lebofsky1986}),
the Near-Earth Thermal Model (NEATM, \cite{Harris1998}), or more complex Thermophysical Models (TPM;
\cite{Lagerros1996,Lagerros1997,Lagerros1998,Rozitis2011}). The STM or NEATM are typically applied to 
survey data \citep[e.g.][]{Tedesco2002,Tedesco2002b,Usui2011,Mainzer2011ApJ736}, while TPM
techniques are used in cases where spin-shape properties are known or for the interpretation of
more complex data sets. But all these techniques have in common that they work for objects with
known orbits where the helio-centric distance, the observer-centric distance and the phase angle are known 
for each individual observing epoch. In this context, the JWST asteroid observations and potential
science cases are presented and discussed by \citet{Norwood2016}, \citet{Rivkin2016} or \citet{Thomas2016},
but the scientific aspects of IR detections of unknown object are not well covered yet.\\

Here, we present JWST-MIRI size and orbit constraints of the outer MBA (10920) 1998~BC1 and a faint, unknown
object (Section~\ref{sec:obs}). For (10920) we complement the MIRI observations by lightcurve
and Gaia DR3 for spin-shape modeling, ATLAS survey data for an estimation of its H-magnitude,
and WISE observations for a radiometric study.
The derived MIRI photometric and astrometric information is given in Section~\ref{sec:miri_results},
followed by a detailed radiometric thermophysical model (TPM) study for (10920) (Section~\ref{sec:radiometry}).
In Section~\ref{sec:stmorbit} we exploit the possibilities and limitations of newly developed STM-ORBIT
method for the determination of the object's physical and orbital properties
just based on IR data alone. We apply the new method first to the known asteroid (10920) for testing
purposes and then to a newly discovered object where no orbit solution is available.
The results are discussed in Section~\ref{sec:discussion} and summarized in Section~\ref{sec:conclusion}.

\section{Observations}
\label{sec:obs}

\subsection{JWST-MIRI observations}
\label{sec:miri_obs}

\begin{figure}[h!tb]
	\resizebox{2.9cm}{!}{\includegraphics{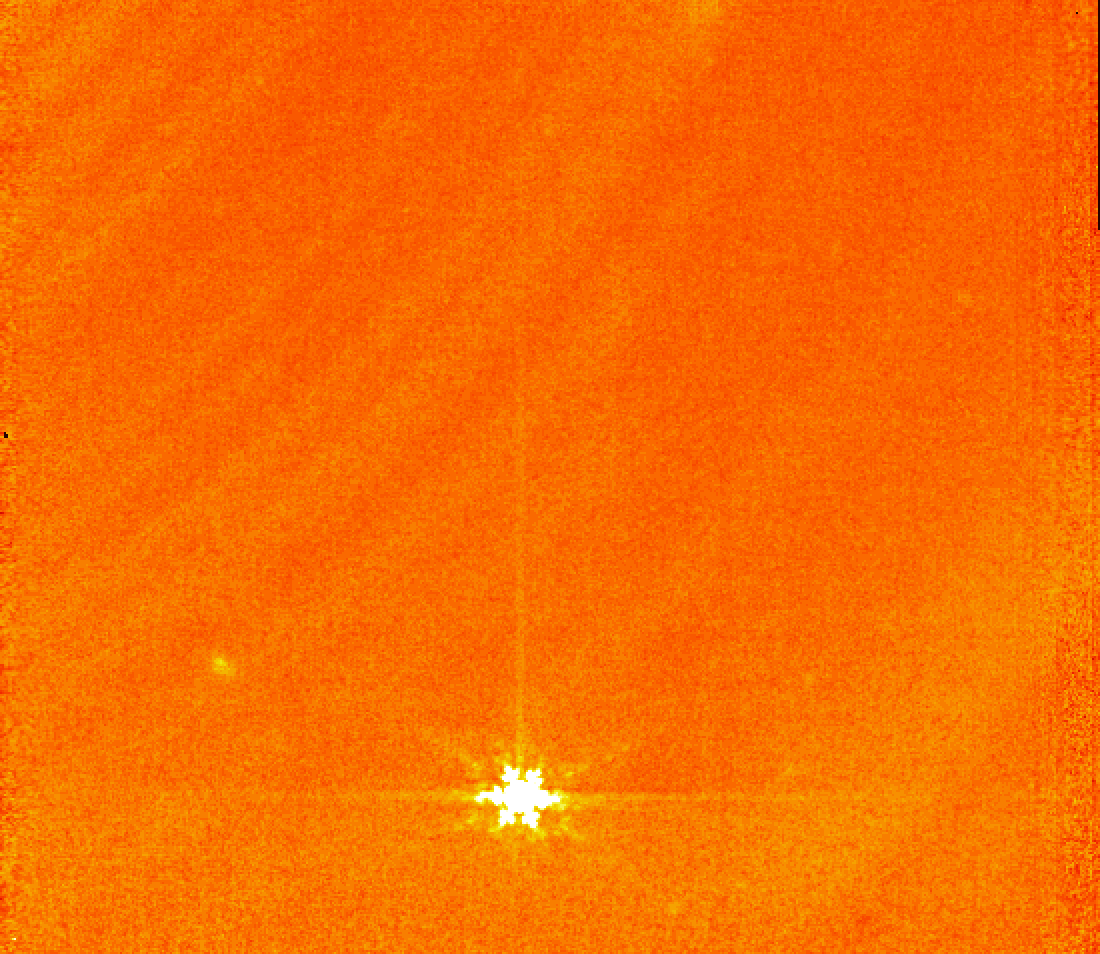}}
	\resizebox{2.9cm}{!}{\includegraphics{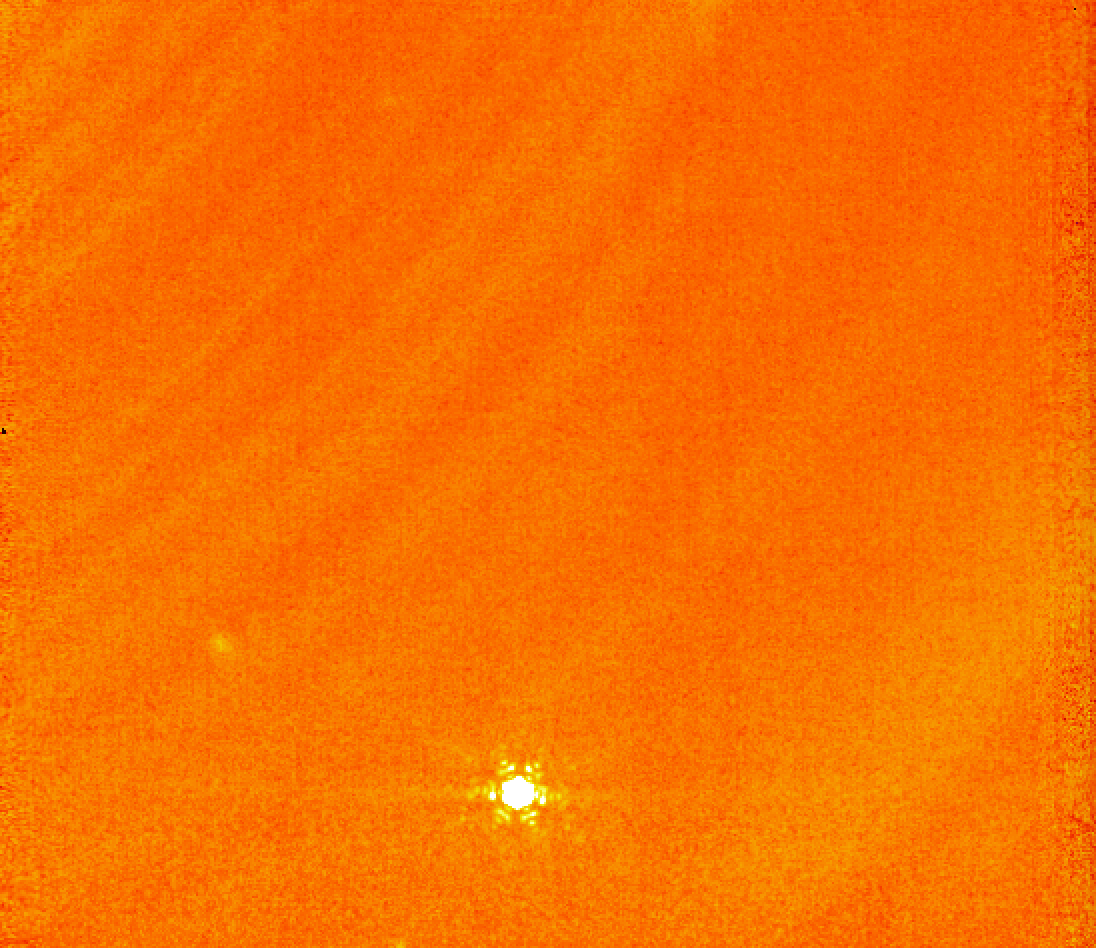}}
	\resizebox{2.9cm}{!}{\includegraphics{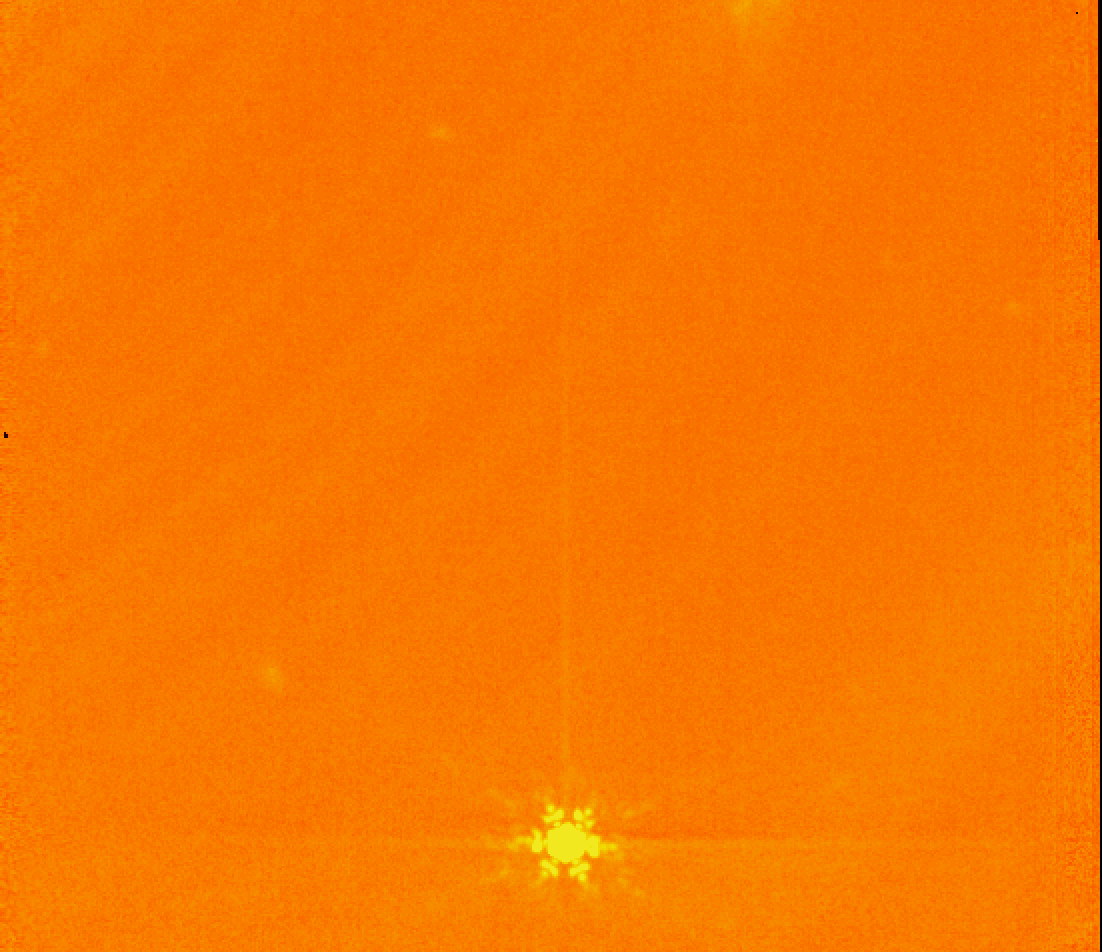}}

        \resizebox{2.9cm}{!}{\includegraphics{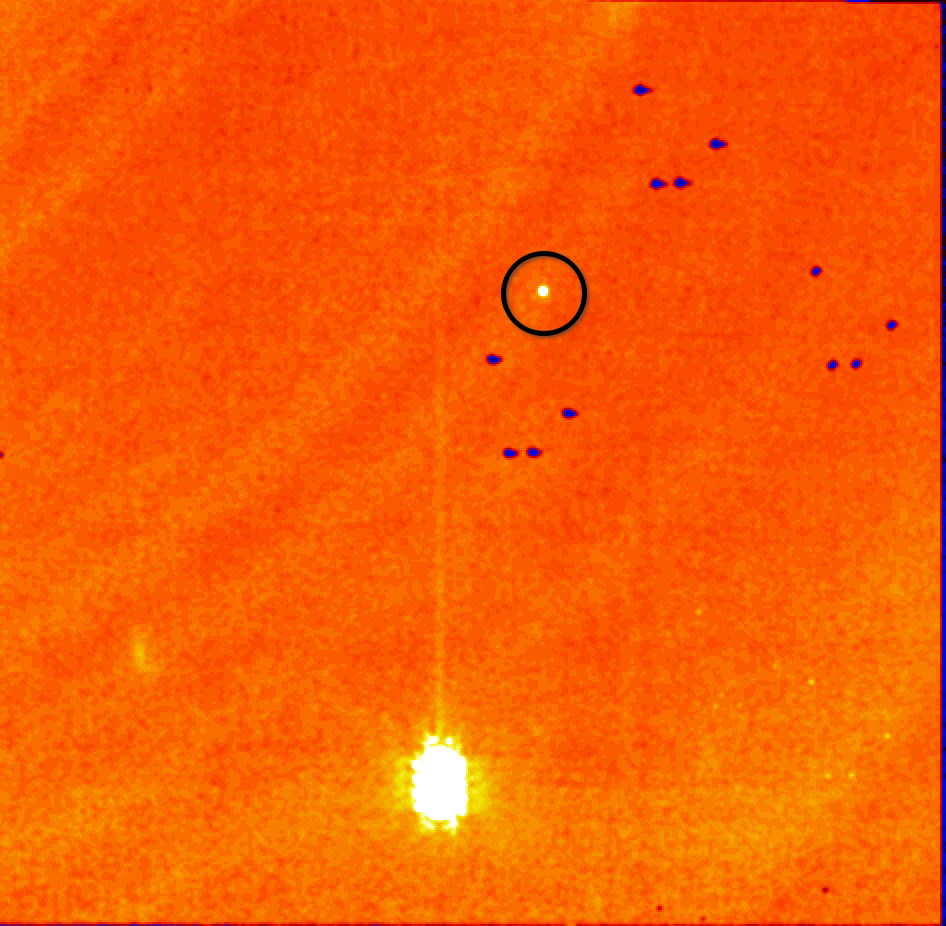}}
        \resizebox{2.9cm}{!}{\includegraphics{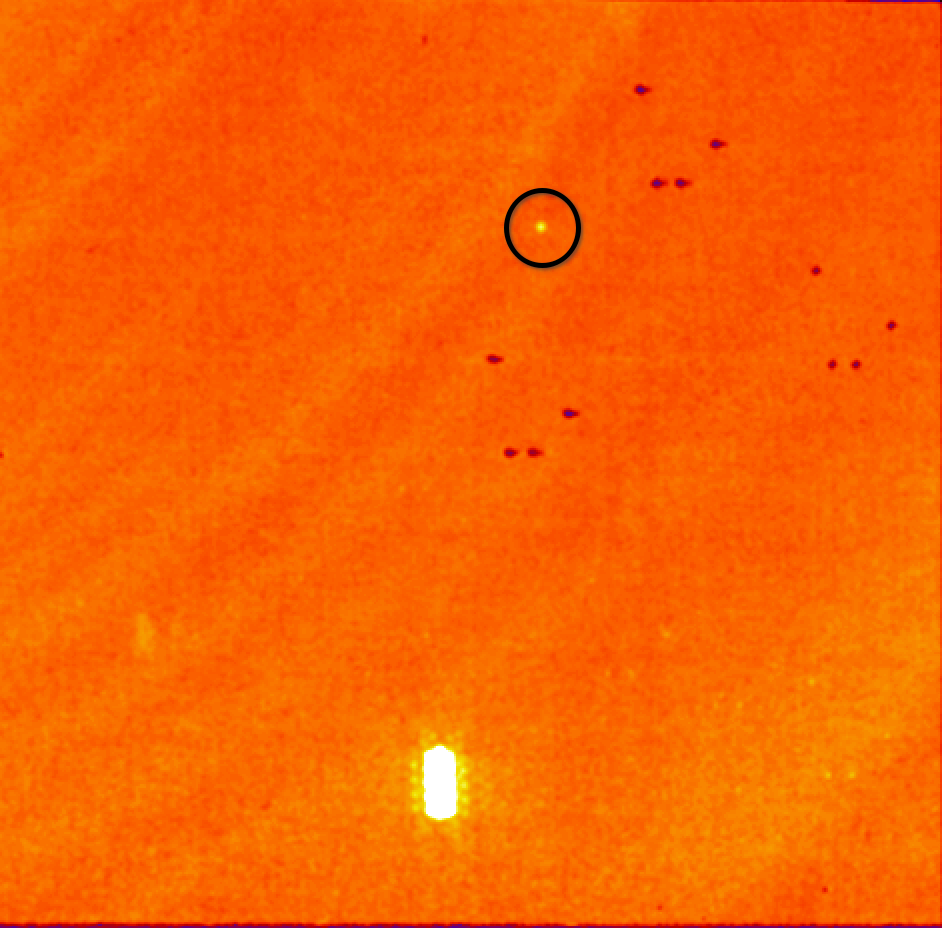}}
        \resizebox{2.9cm}{!}{\includegraphics{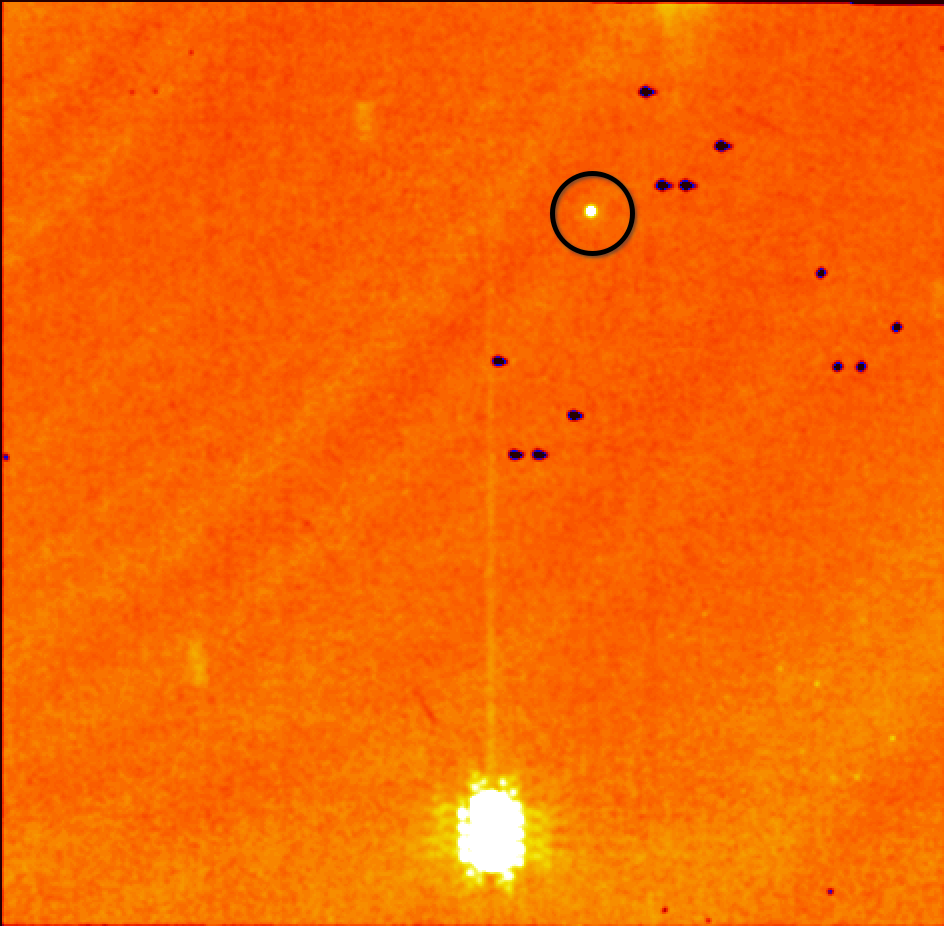}}
%
	\caption{MIRI L3 images of asteroid (10920) (top row) in the F1000W, F1130W, and F1280W band
		 (from left to right), and in the reference frame of (10920). The JWST-centric
		 position of (10920) is 15h41m21.3s in R.A.\, and -19$^{\circ}$12$^{\prime}$17.0$^{\prime \prime}$,
		 the FOV about 50$^{\prime \prime}$ $\times$ 40$^{\prime \prime}$.
		 Bottom row: the corresponding L2 data in the same three bands, but the four
		 dithered images in each band were manually stacked on the new faint asteroid
                 (visible as faint point source in the upper part of the images, marked with a 
         	  black circle). The FOV is about  50$^{\prime \prime}$ $\times$ 50$^{\prime \prime}$.
		  The new object is not visible in the standard L3 data products in the archive.
         \label{fig:newast}}
\end{figure}

A MIRI imaging mode \citep{Bouchet2015} multi-filter measurement sequence
was executed on July 14, 2022
as part of the JWST calibration program "MIRI Imaging Filter Characterization".
The main-belt asteroid (10920) 1998~BC1 was the prime object, however, the individual images also show
a faint object which moved with respect to (10920) and the background sources (see Fig.~\ref{fig:newast}). This faint unknown
object was in the  56.3$^{\prime \prime}$ $\times$ 56.3$^{\prime \prime}$ field-of-view (BRIGHTSKY
sub-array with 512 $\times$ 512 pixels) in all 9 MIRI bands, while the much brighter MBA (10920)
was found to be located at the edge or even outside the FOV\footnote{Field of view} in the long-wavelength measurements.
In each of the nine MIRI filters, a set of 4 dithered images were taken, each with a frame time
of 0.865\,s (FASTR1 readout mode), and an exposure time of 21.632\,s (F0560 band) or 8.653\,s
(for all other bands). 
The JWST data processing is done in three different
stages\footnote{\url{https://jwst-docs.stsci.edu/jwst-science-calibration-pipeline-overview/stages-of-jwst-data-processing}},
using the following pipeline modules for the MIRI imaging observations:
{\tt calwebb\_detector1} (L1): to process raw ramps data into uncalibrated slopes data
(data quality initialization, saturation check, reference pixel correction, jump detection,
slope fitting, reset anomaly correction, first/last frame correction, linearity \&
RSCD\footnote{Reset Switch Charge Decay}  correction, dark subtractions);
{\tt calwebb\_image2} (L2): to process data from uncalibrated slope images into calibrated
slope images (WCS\footnote{World Coordinate System; \url{https://fits.gsfc.nasa.gov/fits_wcs.html}}
information, background subtraction, flat-field correction, flux calibration,
drizzle algorith to produce rectified 2-D products);  {\tt calwebb\_image3} (L3): to process the imaging
data from calibrated slope images to mosaics and source catalogs (refine relative WCS, moving target WCS,
background matching "Skymatch", outlier detection, image combination, source catalog, update of
exposure level products).

For the dedicated (10920) observations presented here, we used the calibrated L2 images 
(corrected for detector and physical effects, and flux calibrated) where
we found 4 individual (dithered) images per filter (for the astrometry and flux extraction).
We also worked with the pipeline-processed calibrated L3 images (see Fig.~\ref{fig:newast} top row)
with the 4 dithered images combined with respect to the moving target (10920) position, 
and manually combined L2 images after stacking onto the new object's position (see Fig.~\ref{fig:newast} bottom row).
For more details on the MIRI imaging mode, see the JWST-specific
documentation\footnote{\url{https://jwst-docs.stsci.edu/jwst-mid-infrared-instrument/miri-observing-modes/miri-imaging}}.
Table~\ref{tbl:miri_obs} summarizes details for the 36 individual data frames.

\begin{table}[h!tb]
  \begin{center}
	  \caption{MIRI imaging mode observations of asteroid (10920) 1998~BC1 were taken
	  as part of a filter characterization calibration program$^{a}$.
          \label{tbl:miri_obs}}
    \begin{tabular}{llllr}
      \hline
      \hline
      \noalign{\smallskip}
	    No.\ & ID$^{b}$ & Date-BEG...END$^{c}$ & Filter$^{d}$ & T$_{exp}$ [s]$^{e}$ \\
      \noalign{\smallskip}
      \hline
      \noalign{\smallskip}
     01       & 2101 & 10:21:08.7...10:21:30.4 &  F560W  &  21.632 \\
     02       & 2101 & 10:24:25.1...10:24:46.8 &  F560W  &  21.632 \\
     03       & 2101 & 10:27:45.9...10:28:07.5 &  F560W  &  21.632 \\
     04       & 2101 & 10:31:05.8...10:31:27.4 &  F560W  &  21.632 \\
     05       & 2103 & 10:36:05.2...10:36:13.8 &  F770W  &   8.653 \\
     06       & 2103 & 10:39:09.5...10:39:18.1 &  F770W  &   8.653 \\
     07       & 2103 & 10:42:13.8...10:42:22.4 &  F770W  &   8.653 \\
     08       & 2103 & 10:45:14.7...10:45:23.3 &  F770W  &   8.653 \\
     09       & 2105 & 10:50:42.6...10:50:51.2 &  F1000W &   8.653 \\
     10       & 2105 & 10:53:46.0...10:53:54.7 &  F1000W &   8.653 \\
     11       & 2105 & 10:56:45.2...10:56:53.8 &  F1000W &   8.653 \\
     12       & 2105 & 10:59:46.0...10:59:54.7 &  F1000W &   8.653 \\
     13       & 2107 & 11:04:14.2...11:04:22.9 &  F1130W &   8.653 \\
     14       & 2107 & 11:07:17.7...11:07:26.3 &  F1130W &   8.653 \\
     15       & 2107 & 11:10:24.6...11:10:33.3 &  F1130W &   8.653 \\
     16       & 2107 & 11:13:26.3...11:13:35.0 &  F1130W &   8.653 \\
     17       & 2109 & 11:17:52.0...11:18:00.6 &  F1280W &   8.653 \\
     18       & 2109 & 11:20:55.4...11:21:04.0 &  F1280W &   8.653 \\
     19       & 2109 & 11:24:00.5...11:24:09.2 &  F1280W &   8.653 \\
     20       & 2109 & 11:27:04.9...11:27:13.5 &  F1280W &   8.653 \\
     21       & 210B & 11:31:38.3...11:31:47.0 &  F1500W &   8.653 \\
     22       & 210B & 11:34:43.5...11:34:52.1 &  F1500W &   8.653 \\
     23       & 210B & 11:37:47.8...11:37:56.5 &  F1500W &   8.653 \\
     24       & 210B & 11:40:51.2...11:40:59.9 &  F1500W &   8.653 \\
     25$^{f}$ & 210D & 11:45:16.9...11:45:25.5 &  F1800W &   8.653 \\
     26       & 210D & 11:48:18.6...11:48:27.2 &  F1800W &   8.653 \\
     27       & 210D & 11:51:26.4...11:51:35.0 &  F1800W &   8.653 \\
     28       & 210D & 11:54:31.5...11:54:40.2 &  F1800W &   8.653 \\
     29$^{g}$ & 210F & 11:58:54.6...11:59:03.3 &  F2100W &   8.653 \\
     30       & 210F & 12:01:58.0...12:02:06.7 &  F2100W &   8.653 \\
     31$^{g}$ & 210F & 12:05:01.5...12:05:10.1 &  F2100W &   8.653 \\
     32$^{g}$ & 210F & 12:08:01.4...12:08:10.1 &  F2100W &   8.653 \\
     33$^{g}$ & 210H & 12:13:11.3...12:13:19.9 &  F2550W &   8.653 \\
     34$^{g}$ & 210H & 12:16:21.6...12:16:30.3 &  F2550W &   8.653 \\
     35$^{g}$ & 210H & 12:19:31.1...12:19:39.8 &  F2550W &   8.653 \\
     36$^{g}$ & 210H & 12:22:35.4...12:22:44.1 &  F2550W &   8.653 \\
      \noalign{\smallskip}
      \hline
    \end{tabular}
	    \\ \vspace*{0.5cm}
  \footnotesize{
   $^{a}$ The calibration proposal ID is 1522. All measurements include also a faint moving source while the prime target
          asteroid was in some cases not in the FOV.
   $^{b}$ The official JWST IDs (all starting with "V01522002001P000000000");
   $^{c}$ the UT start and end times (all taken on 2022-07-14); 
   $^{d}$ the MIRI filter band; 
   $^{e}$ the exposure times.
   $^{f}$ PSF of asteroid (10920) is half outside MIRI image.
   $^{g}$ Asteroid (10920) is outside FOV.}
  \end{center}
\end{table}

The level-2 (L2) products are absolutely calibrated individual exposures. Level-3 (L3)
products have all four dithered images combined (here, stacked on the calculated JWST-centric
position of asteroid (10920) 1998~BC1). These L3 images have effective integration times of
86.528\,s (F560W band) and 34.612\,s (all other bands).
Figure~\ref{fig:newast} shows the F1000W, F1130W, F1280W L3 images (top row) where the dithered
frames were combined (pipeline processed) on the prime target's position. The star-shaped
JWST PSF\footnote{Point-spread function} of asteroid (10920) is dominating the lower quarter
of the frames.
The bottom row shows the same data, but now, the L2 data were manually stacked on
the position of the faint moving object (point source moving up in vertical direction)
which can easily be seen in individual dither frames.

\subsection{Auxiliary observations and lightcurves for (10920)}
\label{sec:aux_obs}

\begin{figure*}[h!tb]
        \resizebox{\hsize}{!}{\includegraphics{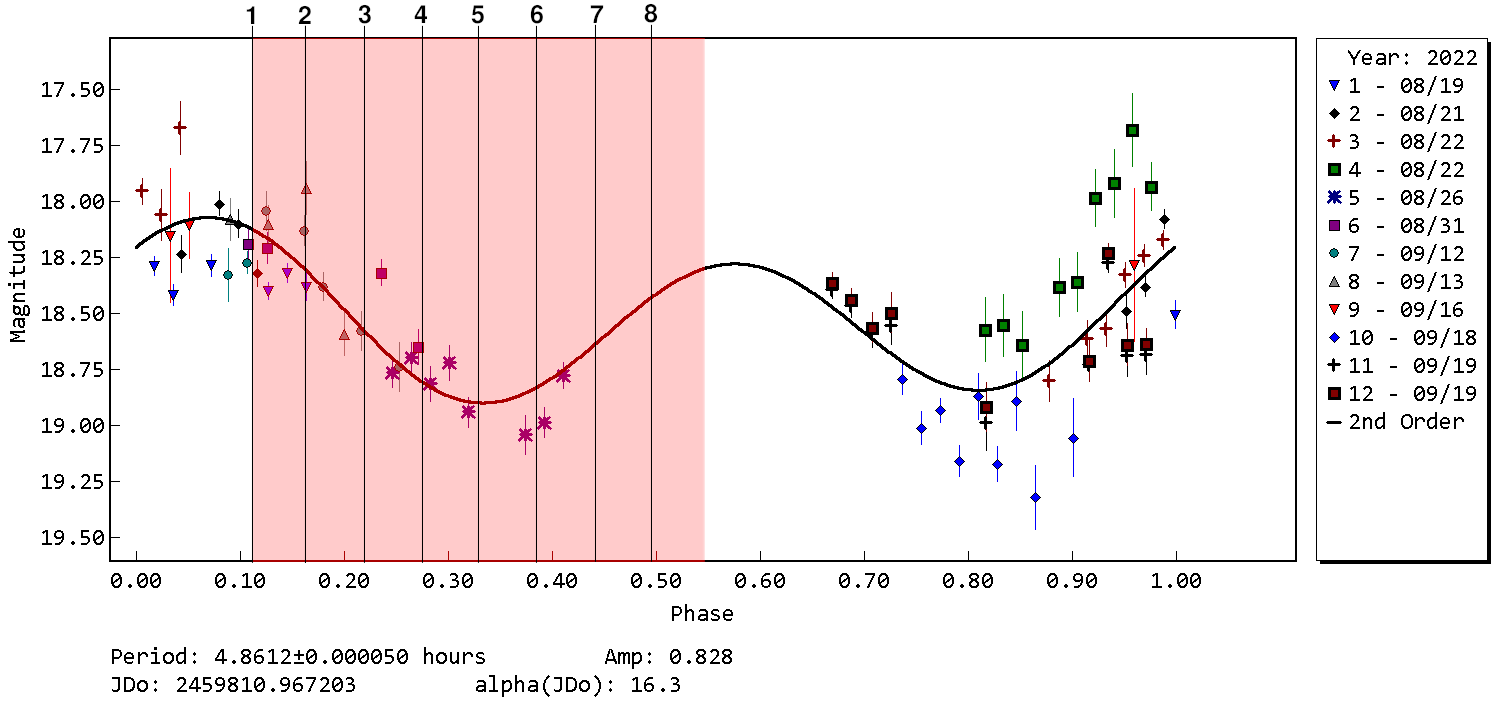}}
        \caption{Lightcurve of (10920) obtained in July/August 2022.
		  Using the period solution, the lightcurve (in the geocentric reference frame)
		  has been phased to the MIRI observation epochs, which are
		  marked with vertical black strokes indicating the beginning
		  of each measurement. The numbers
                  from 1 to 8 correspond to the lines in Table~\ref{tbl:asteroids_miriflx}.
         \label{fig:mba_lc}}
\end{figure*}

Although asteroid (10920) was discovered in 1998, there is very little knowledge about
its physical properties. No shape model or full lightcurves are available, meaning
that the spin state of the body is unknown. Since the asteroid's shape and orientation
are needed to refine the radiometric model, we performed an observation campaign from
August to September 2022 to obtain a lightcurve of (10920) (see Figure~\ref{fig:mba_lc}).
We used the 2 m telescopes of the Las Cumbres Observatory network:
the Faulkes Telescope North (FTN) located at Haleakala Observatory and
the Faulkes Telescope South (FTS) at Siding Spring observatory.
All lightcurve data has been submitted to
LCDB\footnote{Asteroid Lightcurve Data Base at: \url{https://alcdef.org/}}
and can be retrieved by searching for "10920".
Figure~\ref{fig:mba_lc} shows the relevant measurements together with their 
photometric errors. The scatter between the photometric points is not fully compatible 
with the photometric errors. However, the viewing geometry changed slightly over the
32 days of measurements (phase angle range from 16.4$^{\circ}$ to 15.0$^{\circ}$).
Therefore, the scatter on the composite lightcurve for observations taken some weeks apart is
reasonable and expected for this very elongated object.
The lightcurves are close to a sinusoidal function and show a large amplitude
(larger than one magnitude in some cases) and a spin period of about 4.86\,h.
In order to study the spin shape and orientation we used a modified version of the
SAGE modelling technique \citep{Bartczak2018} to fit our lightcurve combined with
the 13 available Gaia DR3\footnote{Gaia Data Release 3,
\url{https://www.cosmos.esa.int/web/gaia/dr3}} sparse photometric measurements
of (10920) \citep{Tanga2022}. We used a simple triaxial ellipsoid shape to fit the data,
since the number of measurements is not enough to study detailed shape features.
However, this simple shape model has proved to work very efficiently to fit
sparse data like the one provided by Gaia \citep{Cellino2015}, and also combined
with ground-based lightcurves \citep{Santana-Ros2015}. We found a solid pole solution
with $\lambda_{ecl} = 178^{\circ}$ and $\beta_{ecl} = 81^{\circ}$ (with an
estimated error of about 20$^{\circ}$), and a very elongated
shape with axis ratios of $a/b = 1.5 - 1.8$ and $b/c = 1.0 - 1.1$. The Gaia and WISE
(see Section~\ref{sec:wise_obs} and Figure~\ref{fig:obsmod_wise}) data are best
matched by a/b $\approx$ 1.5 while groundbased
data pointed to an even more extreme elongation of the object.
The SAGE lightcurve inversion (including the fit to the Gaia DR3 data) also resulted
in a well-determined rotation period of 4.861191 $\pm$ 0.000015\,h. From the spin
solution we can infer that the object is always observed close to an equator-on
viewing geometry, meaning that the aspect angle will not change much from apparition
to apparition. Therefore, changes in the observed cross section of the object are mainly
dominated by the a/b axis ratio while the object is rotating.
We used the derived simple ellipsoidal spin-shape solution to phase the lightcurve
(Fig.~\ref{fig:mba_lc}) back to the epoch of MIRI observations of (10920) (14 July 2022).
The phased lightcurve shows that MIRI observations were obtained mainly during the minima,
implying that most of the measurements were gathered close to the smallest possible
cross section of the body, only the long-wavelength bands were taken at higher 
brightness approaching lightcurve maximum.\\

\begin{figure}[h!tb]
        \resizebox{\hsize}{!}{\includegraphics{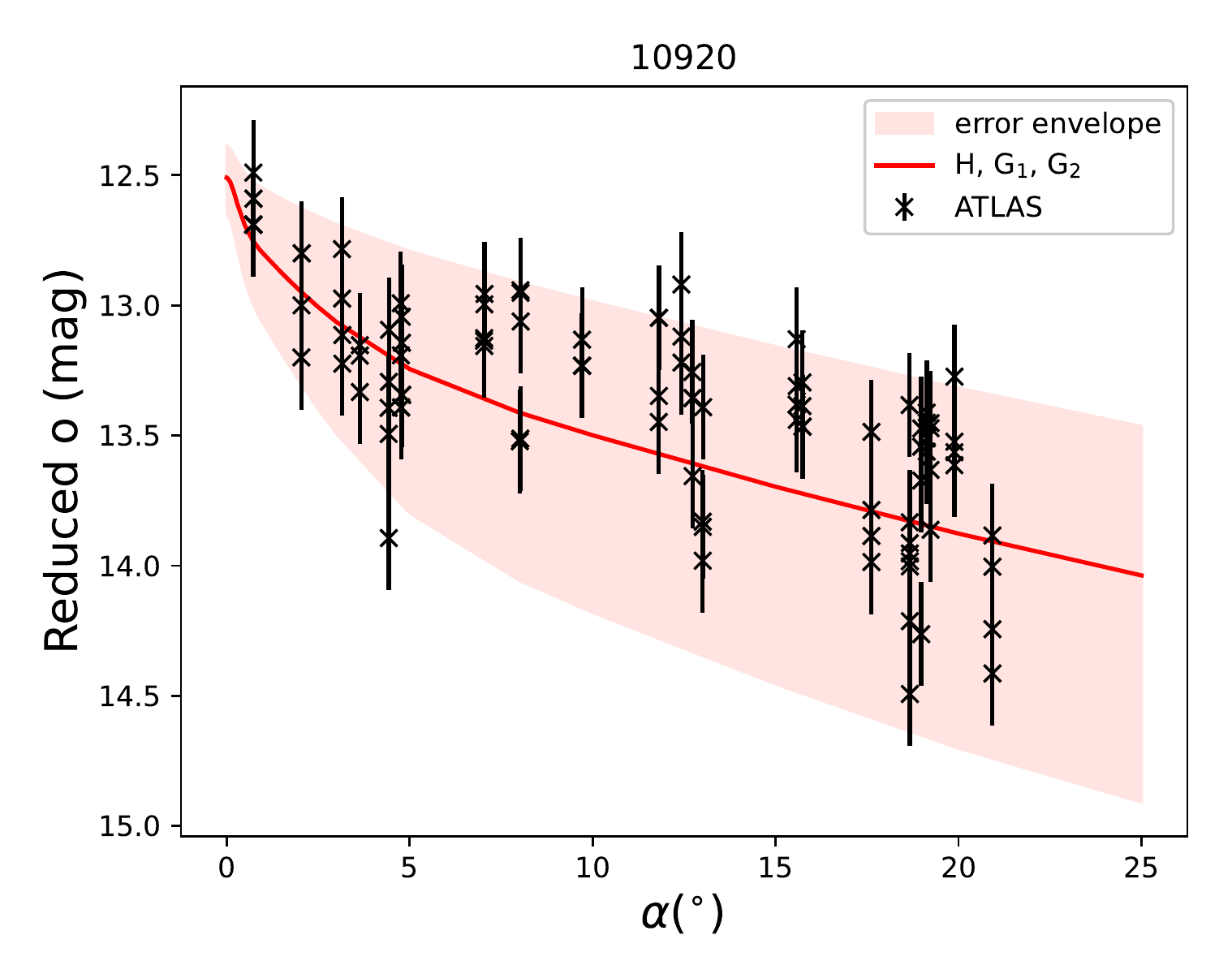}}
        \caption{Fitting the H, G1, G2 photometric phase function to o-band data from
                 the ATLAS survey obtained during three oppositions (2018, 2019/2020, 2021).
                 The single-point uncertainty was assumed to be 0.2\,mag.
         \label{fig:hmag}}
\end{figure}

For the determination of the H-magnitude of (10920)
we fitted the H, G1, G2 photometric phase function of \citet{Muinonen2010} to the data
from the ATLAS survey \citep{Tonry2018}. We utilized only the photometry from the orange filter
as it was more numerous and had a better phase angle coverage than the data from
the cyan filter. The orange passband ($\approx$550 - 820\,nm) largely covers the Johnsons
V band ($\approx$500 - 700\,nm) and derived H values are very similar\footnote{The relation
between the Johnson V-band H$_{V}$ and H$_{o}$ derived from the orange-band ATLAS data is:
H$_V$  =  1.01757 $\pm$ 0.00536 H$_o$ + 0.08286 $\pm$ 0.06104; \citep[][and priv.\ communication for
the difference between cyan and orange-band relations]{Shevchenko2022}.}
Photometry was downloaded through an astroquery wrapper for querying
services at the IAU Minor Planet Center (MPC) \citep{Ginsburg2019}.
Since the aspect changes between the oppositions are minimal, we fitted the data from three
oppositions (2018, 2019/2020, 2021) together (see Fig.~\ref{fig:hmag}).
The fit was performed in the flux domain using a linear least-squares procedure
and we assumed 0.2\,mag photometric uncertainty for the individual data points. Following
\citet{Penttilae2016} we constrained the fits to obtain a physically meaningful solution:
$H_o = 12.51_{-0.12}^{+0.14}$\,mag, and phase-function parameters $G_1=0.27_{-0.17}^{+0.21}$,
$G_2=0.20_{-0.09}^{+0.08}$. This compares very well with values derived by
\citet{Mahlke2021} and a previous study by \citet{Oszkiewicz2011}.


\subsection{Auxiliary WISE observations for (10920)}
\label{sec:wise_obs}


The Wide-field Infrared Survey Explorer \citep[WISE,][]{Wright2010} mapped
in 2010 large parts of the sky in four IR bands at 3.4, 4.6, 12, and 22\,$\mu$m (W1-W4).
After the cryo-phase, the mission was continued as Near-Earth Object WISE
\citep[NEOWISE,][]{Mainzer2014}, but only in the short-wavelength channels W1 and W2.
\citet{Masiero2011} published a size-albedo solution for (10920) based on multiple
W3 (15x) and W4 (16x) detections.
They found a size of 14.436 $\pm$ 0.267\,km and a geometric albedo of
p$_V$=0.0847 $\pm$0.0159 (based on H = 12.50\,mag).
The fitted beaming parameter $\eta$ is given with 1.058 $\pm$ 0.038.
\citet{Nugent2015} published a size of 11.12 $\pm$ 2.85\,km and p$_V$ = 0.11 $\pm$ 0.07 (H=12.80\,mag)
as their best solution (via NEATM $\eta$-fit with $\eta$ = 0.95 $\pm$ 0.17).
Detections from W1 (6x) and W2 (6x) were used for these calculations. They also see a 0.31\,mag amplitude
of the WISE 4.6\,$\mu$m (W2) lightcurve, however, based on 6 data points only.
\citet{Mainzer2019} list another solution where the fit was done on
W2 (9x), W3 (9x), and W4 (10x) detections:
D=15.712 $\pm$ 0.187\,km, p$_V$ = 0.082 $\pm$0.012, $\eta$ = 1.087 $\pm$ 0.022.
All WISE-specific radiometric solutions are based on the Near-Earth Asteroid Thermal Model
\citep[via NEATM,][]{Harris1998} which uses the beaming parameter $\eta$ to obtain the best
fit to the observed spectral slope. $\eta$ depends on the object's rotation, thermal, surface,
and emissivity properties, and is also influenced by the observed wavelength regime and 
observing geometry (heliocentric distance and phase angle). The NEATM is closely connected to
the Standard Thermal Model \citep[STM,][]{Lebofsky1986} which uses a fixed $\eta$ of 0.756,
and which is widely applied to main-belt asteroids \citep[e.g.][]{Tedesco2002,Tedesco2002b,Delbo2015}.

For our radiometric study of asteroid (10920), we extracted all W2, W3, and W4 WISE measurements
with photometry quality flags 'A' (SNR$\ge$10) or 'B' (3$\le$SNR$<$10), avoiding saturation, moon separation
of less than 15$^{\circ}$ or close-by background sources. The extracted magnitudes
have been converted to fluxes \citep[see][]{Wright2010} and color-corrected with correction
factors of 1.23, 0.97, and 0.98 for W2, W3, and W4, respectively\footnote{The color correction allows
to produce mono-chromatic flux densities at the WISE reference wavelengths. These corrections are
based on difference between the asteroid's spectral shape compared to Vega reference spectrum which
was used for establishing the WISE photometric calibration.}.
For the WISE absolute flux errors we considered the observational flux errors, an estimated
error for the color correction and absolute flux errors which added up to minimum values of
15, 7, and 7\% in the W2, W3, and W4 bands, respectively. The full list of WISE fluxes and errors are
given in Table~\ref{app:wise_fluxes}.

\section{MIRI-related results}
\label{sec:miri_results}
\subsection{Astrometry}

We determined the PSF centroid of asteroid (10920) (28 L2 images) and the newly discovered object (all 36 L2 images)
by a simple 2-D Gauss fit, and used the WCS header information to translate the pixel coordinates
into a rough estimate of their R.A.\ and Declination (see Table~\ref{tbl:astrometry}).
The WCS-translated source positions gave a total apparent sky path of
7.27$^{\prime \prime}$ in 1:40:42.82\,h or an apparent motion
of 4.33$^{\prime \prime}$/h, in excellent agreement with JPL/Horizons predictions.

However, the absolute coordinates for (10920) are in poor agreement with the
true JWST-centric ephemeris
of the object, as determined by JPL/Horizons and our own orbital calculation. They show
an offset of about 0.17$^{\prime \prime}$, which corresponds to about 1.5 pixels of
the MIRI detector. The offset is not aligned with the direction of motion, and
therefore not due to a timing issue.

In order to track down the source of this issue we manually re-measured
the position of the asteroid in one of the shorter wavelength images,
using stars that are visible in the IR image while still having
counterparts in the optical. Unfortunately, only 2 Gaia DR3 sources
are contained in the FoV of the MIRI frames. However, we located a
deep optical image of the same area of the sky obtained by the Dark Energy
Camera (DECam) \citep{Flaugher2015} in their public archive\footnote{\url{https://www.darkenergysurvey.org/the-des-project/data-access/}}.
The frame, dated 2019 February 23, has an exposure time of 56\,s,
and contains a significant number of optical sources falling close to
the MIRI sky-print.
We carefully measured the position of 44 such sources versus a
Gaia-based solution of the entire DECam chip, and used these positions
to create a secondary catalog for applying to the astrometry of the MIRI
frame. A total of 16 of such sources showed a detectable counterpart
in the MIRI exposure, and could be used to perform a full astrometric
solution on it, completely independent from the WCS coefficients. This
solution evidenced a bias that exactly compensates the offset observed
in the WCS-based astrometry. An astrometric measurement of asteroid
(10920) with respect to this self-derived solution fits the existing
ephemeris of the asteroid with residuals of $<$0.01$^{\prime \prime}$
in both coordinates (and has a formal uncertainty of roughly
$\pm$0.03$^{\prime \prime}$).
This test proves that the bias we observed in the WCS-based astrometry
is due to the WCS solution present in the MIRI images, and it is not a
manifestation of a physical offset in the object's true sky position,
nor of a problem with the ephemeris of the observing spacecraft.

\begin{table*}[h!tb]
  \begin{center}
  \caption{The WCS-based R.A.\ and Dec.\ coordinates of (10920) and the new object extracted from
   L2 products via Gaussian centroid fitting to the sources. MJD is the observation mid-time (UT).
          \label{tbl:astrometry}}
    \begin{tabular}{lrrrrr}
      \hline
      \hline
      \noalign{\smallskip}
                &              & \multicolumn{2}{c}{(10920) 1998~BC1} & \multicolumn{2}{c}{new object} \\
       No.\     & MJD          & R.A.\ [deg] & Dec.\ [deg] & R.A.\ [deg] & Dec.\ [deg] \\
      \noalign{\smallskip}
      \hline
      \noalign{\smallskip}
       01 & 59774.431476 & 235.3396411 & -19.2048876 & 235.3453320 & -19.2049123  \\
       02 & 59774.433749 & 235.3395726 & -19.2048763 & 235.3454998 & -19.2049910  \\
       03 & 59774.436073 & 235.3395027 & -19.2048649 & 235.3456712 & -19.2050717  \\
       04 & 59774.438387 & 235.3394330 & -19.2048534 & 235.3458419 & -19.2051356  \\
       05 & 59774.441776 & 235.3393309 & -19.2048367 & 235.3460921 & -19.2052401  \\
       06 & 59774.443910 & 235.3392667 & -19.2048262 & 235.3462495 & -19.2053028  \\
       07 & 59774.446043 & 235.3392024 & -19.2048156 & 235.3464069 & -19.2053612  \\
       08 & 59774.448136 & 235.3391394 & -19.2048053 & 235.3465613 & -19.2054410  \\
       09 & 59774.451932 & 235.3390251 & -19.2047865 & 235.3468414 & -19.2055228  \\
       10 & 59774.454055 & 235.3389612 & -19.2047760 & 235.3469981 & -19.2055865  \\
       11 & 59774.456128 & 235.3388987 & -19.2047658 & 235.3471510 & -19.2056832  \\
       12 & 59774.458222 & 235.3388357 & -19.2047555 & 235.3473055 & -19.2057778  \\
       13 & 59774.461326 & 235.3387422 & -19.2047401 & 235.3475346 & -19.2058274  \\
       14 & 59774.463449 & 235.3386783 & -19.2047296 & 235.3476912 & -19.2059116  \\
       15 & 59774.465613 & 235.3386131 & -19.2047190 & 235.3478509 & -19.2059585  \\
       16 & 59774.467716 & 235.3385498 & -19.2047086 & 235.3480061 & -19.2060255  \\
       17 & 59774.470790 & 235.3384572 & -19.2046934 & 235.3482329 & -19.2061170  \\
       18 & 59774.472913 & 235.3383933 & -19.2046829 & 235.3483896 & -19.2061859  \\
       19 & 59774.475056 & 235.3383288 & -19.2046723 & 235.3485477 & -19.2062493  \\
       20 & 59774.477190 & 235.3382645 & -19.2046618 & 235.3487051 & -19.2063141  \\
       21 & 59774.480355 & 235.3381692 & -19.2046462 & 235.3489387 & -19.2064240  \\
       22 & 59774.482498 & 235.3381047 & -19.2046356 & 235.3490968 & -19.2064844  \\
       23 & 59774.484631 & 235.3380404 & -19.2046250 & 235.3492542 & -19.2065476  \\
       24 & 59774.486754 & 235.3379765 & -19.2046145 & 235.3494109 & -19.2066109  \\
       25$^{a}$ & 59774.489829 & ---   & ---         & 235.3496377 & -19.2067233  \\
       26 & 59774.491932 & 235.3378206 & -19.2045890 & 235.3497929 & -19.2067603  \\
       27 & 59774.494105 & 235.3377552 & -19.2045782 & 235.3499533 & -19.2068342  \\
       28 & 59774.496248 & 235.3376906 & -19.2045677 & 235.3501114 & -19.2069013  \\
       29$^{b}$ & 59774.499293 & ---   & ---         & 235.3503361 & -19.2069632  \\
       30 & 59774.501416 & 235.3375350 & -19.2045421 & 235.3504927 & -19.2070349  \\
       31$^{b}$ & 59774.503539 & ---   & ---	     & 235.3506494 & -19.2070889  \\
       32$^{b}$ & 59774.505622 & ---   & ---	     & 235.3508031 & -19.2071688  \\
       33$^{b}$ & 59774.509208 & ---   & ---	     & 235.3510677 & -19.2073011  \\
       34$^{b}$ & 59774.511411 & ---   & ---	     & 235.3512302 & -19.2073613  \\
       35$^{b}$ & 59774.513604 & ---   & ---	     & 235.3513921 & -19.2074352  \\
       36$^{b}$ & 59774.515738 & ---   & ---	     & 235.3515495 & -19.2074973  \\
      \noalign{\smallskip}
      \hline
    \end{tabular}
            \\
  \footnotesize{
	  $^{a}$ Asteroid (10920) at edge, no centroid position possible.
	  $^{b}$ Asteroid (10920) is outside FOV.}
  \end{center}
\end{table*}

The new object was detected in all 36 L2 frames (9 filters with 4 dithered frames in each band).
The WCS-translated source positions are also given in Table~\ref{tbl:astrometry}.
The new target moved by 23.00$^{\prime \prime}$ during the 2:01:20.24 hours, corresponding to an apparent
motion of 11.37$^{\prime \prime}$/hour, or about 2.6 times faster than the outer main-belt asteroid (10920).
We also determined the absolute astrometric solution for the new object based on the DECam deep optical
image and connected to a Gaia-based solution, as described above for (10920). This allowed us to determine
highly accurate ($<$ $\pm$0.05$^{\prime \prime}$) positions for the new object in the short-wavelength
MIRI bands (where the stars are still visible). The observed arc is in principle too short for the 
MPC\footnote{\url{https://www.minorplanetcenter.net/}} to designate it (it will be kept in the
unpublished and unreferenced "Isolated Tracklet File"),
but future projects like LSST\footnote{\url{https://www.lsst.org/}} or the NEO Surveyor\footnote{\url{https://neos.arizona.edu}}
might be able to pick it up again. The JWST positions of this new object from July 2022 will then be very
useful for orbit calculations.

\subsection{IR Photometry}

Before we worked on the flux extraction for both moving objects, we looked at all calibration
stars\footnote{HD\,2811 (sub-array mode: SUB64), HD\,163466 (SUB64), HD\,180609 (SUB128),
2MASS~J17430448+6655015 (FULL), 2MASS~J18022716+6043356 (BRIGHTSKY), and BD+60-1753 (SUB256)}
which were observed as part of the MIRI photometric calibration
program\footnote{JWST Calibration program CAL/CROSS 1536}.
They were taken in the same MIRI imaging mode, the same readout (FASTR1) and dither mode
(4-POINT-SETS), and only the sub-array settings (SUB64, SUB128, SUB256, BRIGHTSKY, FULL)
and the exposure times were different. These stars cover the MIRI flux range
between 0.05\,mJy (in F2550W) and above 300\,mJy (in F560W), similar to the flux levels
of the asteroid (10920). We applied aperture photometry (up to a point where the growth curve
flattened out) with the sky background subtracted (calculated within an annulus at sufficient
distance from the star, avoiding background sources and image artifacts).
When comparing these aperture fluxes with the corresponding
model fluxes\footnote{\url{https://www.stsci.edu/hst/instrumentation/reference-data-for-calibration-and-tools/astronomical-catalogs/calspec}}
\citep{Gordon2022}, we found an agreement of typically 5\% or better, but with some
outliers on the 10-15\% level, mostly in cases where the star was close to the edge 
of the FOV or near a bright artifact (visible in level-2 products). Only for HD\,2811,
we found MIRI fluxes (level-2 and level-3 products) which are about 1.5-2.2 times higher
than the model fluxes. The reason for the HD\,2811 discrepancy is not clear, but \citet{Rieke2022}
flagged this star as less reliable due to obscuration (A$_V$ $>$ 0.2\,mag). We
excluded this star from our study.

Following the experience from our stellar calibration analysis, we performed similar aperture photometry
for the two moving objects, both on L2 and L3 data (or manually stacked level-2 frames for the new object).
For the flux error we considered the scatter between the individual L2 fluxes and the SNRs of the combined
images. In case of (10920), the SNRs are well above 100 in all cases and the scatter between fluxes derived
from the four L2 frames agree within 5\%. Therefore, we took 5\% as the measurement error, except
for the single F2100W measurement where a significant part of the source PSF is outside the FOV.
Here, we performed a "half-source photometry" (multiplied by 2) and we estimated a 10\% measurement error.

For the new object the fluxes are much lower (note, that we use mJy for (10920) and $\mu$Jy for the
new object!). Still, the SNR for this object ranges between 10 and 20 (individual L2 frames) and
goes to 25 in some of the final stacked images. At longer wavelengths, the object increases in brightness,
but also the background level goes up, and the detection was more difficult.

\begin{figure}[h!tb]
        \resizebox{\hsize}{!}{\includegraphics{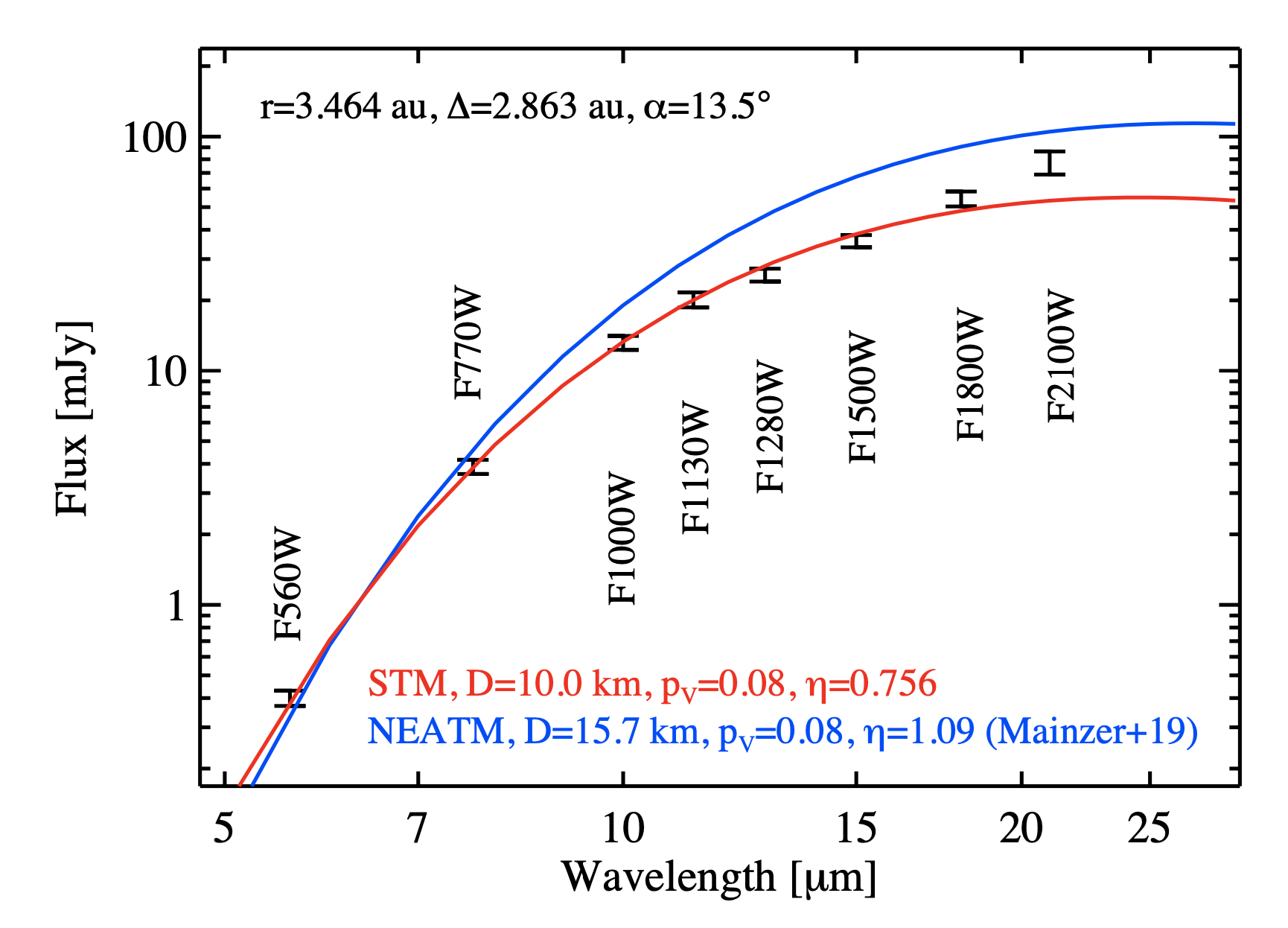}}
	\caption{The extracted MIRI fluxes for asteroid (10920), shown together with 
	         two different model predictions for the JWST observing epoch and geometry.
		 The blue line shows the NEATM prediction, using the \citep{Mainzer2019}
		 solution, the red line is based on a STM prediction, but using a 10 km
		 diameter for (10920).
         \label{fig:mba_sed}}
\end{figure}

\begin{figure}[h!tb]
        \resizebox{\hsize}{!}{\includegraphics{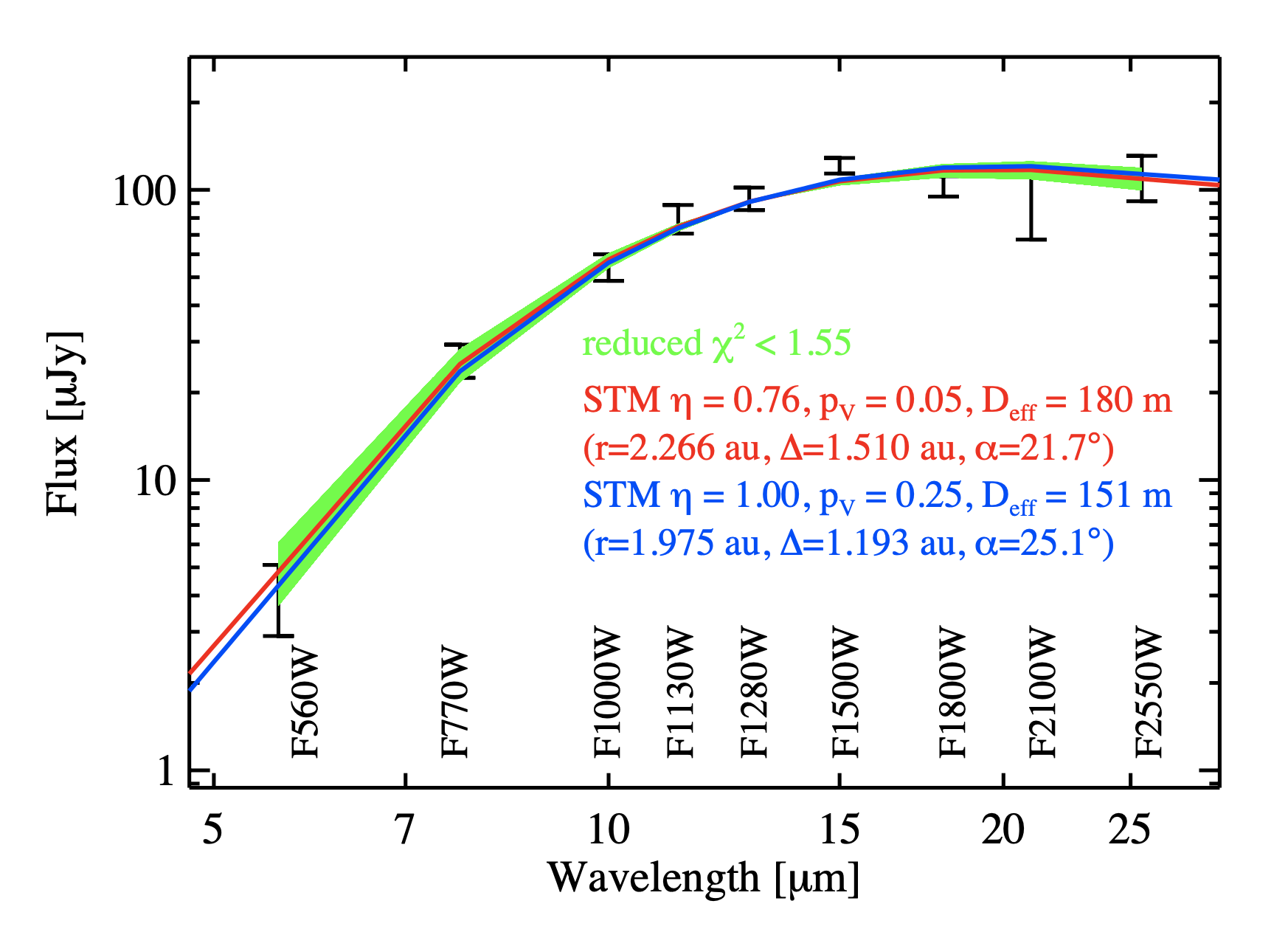}}
        \caption{The extracted MIRI fluxes for the new object, shown together with
                 STM predictions for two different $\eta$-values which correspond
                 also to different best-fit $r, \Delta, \alpha$ observing geometries.
                 The red line shows the $\eta = 0.756$ (p$_V$ = 0.05, D$_{eff}$ = km),
                 the blue line the $\eta = 1.0$ (p$_V$ = 0.25, D$_{eff}$ = km)
		 STM solution, and the green envelope all orbit solutions with
		 $\chi^2$ $<$ 1.55. Note that the flux scale is given in $\mu$Jy.
         \label{fig:newast_sed}}
\end{figure}

The MIRI flux
calibration is based on the assumption of  ${F}_{\mathrm{ref}}(\lambda )=\mathrm{const.}$ \citep{Gordon2022}.
It is therefore necessary to color-correct the extracted in-band fluxes to obtain mono-chromatic flux densities
at the MIRI reference wavelengths. Based on the MIRI filter transmission curves and a WISE-based model spectrum
for (10920) \citep{Mainzer2019}, we calculated corrections factors of 1.15, 1.02, 0.99, 1.00, 0.99, 0.99, 1.00,
1.01, and 1.01 in the 9 MIRI bands from F560W to F2550W (roughly corresponding to the corrections for
$\approx$200-240\,K black bodies).
For the color correction we assume a 2\% error (5\% in the F560W band), and for the MIRI absolute
flux calibration another 5\% error in all bands. All errors were added quadratically.
The results are presented in table~\ref{tbl:asteroids_miriflx} and in Figures~\ref{fig:mba_sed} and \ref{fig:newast_sed}.
Note that these multi-band
fluxes cannot be used to see rotational (lightcurve) flux variations directly as the flux changes
are dominated by the change in wavelength. Only in combination with good-quality spin-shape
model solutions there is a possibility to (partially) separate rotational from spectral
flux variations (see discussion for (10920) in Section~\ref{sec:radiometry}).


\begin{table*}[h!tb]
  \begin{center}
	  \caption{Extracted MIRI fluxes for the main-belt asteroid (10920)$^{a}$ (in mJy) and the new faint asteroid$^{b}$ (in $\mu$Jy).
	   \label{tbl:asteroids_miriflx}}.
  \begin{tabular}{r|rr|rrl|rrl}
      \hline
      \hline
      \noalign{\smallskip}
	    &     &         & \multicolumn{3}{c}{MBA (10920)}       & \multicolumn{3}{c}{New Object} \\
	  No. & MJD & $\lambda$ [$\mu$m]   & flux [mJy] & error$^{c}$ [mJy] & Comments & flux [$\mu$Jy] & error$^{c}$ [$\mu$Jy] & Comments \\
      \noalign{\smallskip}
      \hline
      \noalign{\smallskip}
	1 & 59774.434921 &  5.60 &  0.40 & 0.03 & L2 1-4   \& L3 &   4.0 &  1.1 & low SNR \\
	2 & 59774.444966 &  7.70 &  3.89 & 0.27 & L2 5-8   \& L3 &  25.9 &  3.4 & \\ 
	3 & 59774.455084 & 10.00 & 13.19 & 0.90 & L2 9-12  \& L3 &  54.3 &  5.7 & \\ 
	4 & 59774.464526 & 11.30 & 20.14 & 1.48 & L2 13-16 \& L3 &  79.7 &  9.0 & \\ 
	5 & 59774.473988 & 12.80 & 25.67 & 1.63 & L2 17-20 \& L3 &  93.5 &  8.3 & \\ 
	6 & 59774.483560 & 15.00 & 35.83 & 2.15 & L2 21-24 \& L3 & 121.3 &  7.5 & \\ 
	7 & 59774.493029 & 18.00 & 54.31 & 3.98 & L2 26-28 \& L3 & 103.1 &  8.3 & high bgr\\ 
	8 & 59774.502468 & 21.00 & 77.63 & 8.79 & L2 30    \& L3 &  91.8 & 24.4 & high bgr\\
	9 & 59774.512490 & 25.50 & ---   & ---  & out of FOV     & 111.2 & 19.8 & high bgr\\
      \noalign{\smallskip}
      \hline
    \end{tabular}
            \\
  \footnotesize{
   $^{a}$ The asteroid (10920) was at a helio-centric distance of r=3.5640\,au, a JWST-centric distance
          of 2.86328\,au, a solar elongation of 125.87$^{\circ}$, and seen under a
          phase angle $\alpha$=13.51$^{\circ}$ at observation mid-time (2022-Jul-14 11:20 UT).
   $^{b}$ For the faint new asteroid, only its positions over two hours and the apparent motion
          are known (see Table~\ref{tbl:astrometry}). Both objects had a solar elongation
          ($\lambda$ - $\lambda_{sun}$) = 125.8$^{\circ}$, as seen from JWST.
   $^{c}$ The flux errors are standard deviations of the photometry of the four level-2 images,
          combined with estimated errors for the color correction and absolute flux calibration.}
  \end{center}
\end{table*}

\section{Radiometric study for MBA (10920)}
\label{sec:radiometry}

Figure~\ref{fig:mba_sed} shows our calibrated MIRI fluxes of MBA (10920) together with two
flux predictions: (1) A NEATM prediction (blue line) for this specific JWST-centric observing geometry
on July 14, 2022. The model parameters (D, p$_V$, and $\eta$) were taken from \citet{Mainzer2019}.
(2) In addition, we calculated a STM prediction ($\eta$=0.756) for a 10 km diameter sphere (with p$_V$=0.082
as in the NEATM calculations) shown in red. The NEATM fluxes are too high (by almost a factor of 2 at 15\,$\mu$m)
while the STM fluxes are in nice agreement in the wavelength range up to 18\,$\mu$m. Only the 21 $\mu$m point
is off, but it was taken when the source was at the edge of the MIRI array. And, as can be seen from
Figure~\ref{fig:mba_lc}, the 21 $\mu$m point (No.\ 8 in Table~\ref{tbl:asteroids_miriflx}) falls close to the lightcurve
maximum which might also explain the discrepancy with the simple model prediction.

The 8 MIRI fluxes (or only 7 as the 21-$\mu$m flux is uncertain) span a wide wavelength range close
to the object's thermal emission peak. We fitted these data with the STM by using different beaming
values (and albedos). A beaming parameter of $\eta$ = 0.76 ($\pm$ 0.03) (or 0.73 $\pm$ 0.04 when using 7 fluxes only)
produces the best-fit to the observations. In conclusion, the STM assumption with $\eta$=0.756 is working very well in
case of our MIRI observations of (10920). The corresponding STM radiometric size is 10.0 $\pm$ 0.2\,km,
the geometric albedo 0.13 (assuming an absolute magnitude H = 12.8\,mag).

The WISE data were taken in mid January 2010 (11 detections in W3/W4) at a phase angle of
+18.8$^{\circ}$ (trailing the Sun) and in early July 2010 (14 detections in W3/W4) at a phase angle of 
-17.9$^{\circ}$ (leading the Sun). These dual-band and dual-epoch data are best fit for a pro-grade
spin of (10920) with a thermal inertia in the range between 9 and 35\,J\,m$^{-2}$s$^{-0.5}$K$^{-1}$,
with the best-fit value of 15\,J\,m$^{-2}$s$^{-0.5}$K$^{-1}$.
The WISE flux variations point to an elongated body with an estimated axis ratio a/b $\approx$ 1.5
(see Fig.~\ref{fig:obsmod_wise}). However, a 
simple ellipsoidal shape cannot fully explain the observed thermal lightcurves. Instead, the variations
might point to a (contact-)binary system. The relatively high thermal inertia (as compared to other
outer main-belt objects, see \cite{Delbo2015}) leads to a radiometric size of 14.5 - 16.5\,km (size of
an equal-volume sphere), larger than our STM predictions and close to previously published WISE solutions.
The best-fit radiometric size for a spherical shape model is 14.7\,km, the best-fit for the ellipsoidal
shape is a bit larger at 15.3\,km. Based on the WISE-W3 and W4 data (Fig.~\ref{fig:obsmod_wise}),
we estimated a minimum cross section of (10920) of about 10.5\,km (similar to our initial STM fit
to the JWST data in Fig.~\ref{fig:mba_sed}), and a maximum cross-section of 18\,km.
However, the MIRI observations were taken close to the lightcurve minimum (see Fig.~\ref{fig:mba_lc})
and a radiometric TPM analysis of these data alone would result in a smaller
cross-section (about 11-13\,km in diameter).
Adjusting the MIRI fluxes via our optical lightcurve data to a lightcurve-median value 
(followed by a radiometric study using a spherical-shape model) is not easily possible.
Lightcurves are changing with phase angle and, the interpretation in terms of 
spin-shape properties depends on surface scattering models \citep{Lu2019}.
The IR fluxes, on the other hand, are influenced in a wavelength-dependent fashion by thermally-relevant
properties, like the surface roughness or thermal inertia \citep{Mueller2002}. Only at longer wavelengths,
close to the thermal emission peak and beyond, the thermal fluxes are less sensitive
to thermal properties and follow closely the object's changes in cross section.
A simple (optical) lightcurve-based correction of the MIRI fluxes is therefore not possible.

The object's albedo is connected to the absolute magnitude. H = 12.5\,mag 
corresponds to p$_V$ = 0.07, as expected for a typical object in the outer main-belt region. A smaller
(larger) H-magnitude would lead to a higher (lower) albedo and p$_V$ values between 0.05 and 0.09 are
compatible with the available absolute magnitude fits (see Fig.~\ref{fig:hmag}).


\section{STM-ORBIT method}
\label{sec:stmorbit}

In this section, we assume that the orbits of both asteroids are not known. The goal is to 
constrain the size, helio-centric distance (at the same time the JWST-centric distance $\Delta$
and the phase angle $\alpha$), and possibly also their orbital 
parameters by just using the MIRI fluxes, the JWST-centric R.A.\ and Dec.\ coordinates,
the derived apparent motion, and the solar-elongation of the targets at the time of the
MIRI measurements.

\subsection{Orbit calculations}

For the MBA (10920) we calculated more than 9300 orbits that are compatible with the observed
JWST-centric RA/Dec and motion direction of the object, the apparent solar elongation,
and the specific apparent motion of 4.33$^{\prime \prime}$/hour.
The calculations were done via a ranging approach \citep{Virtanen2001,Oszkiewicz2009} using the "Find\_Orb"
software\footnote{\url{https://www.projectpluto.com/find_orb.htm}}.
Similar ranging computations have been used before, but usually in the context of unconstrained
systematic ranging concepts for the orbit determination of near-Earth asteroids, for impact
probability calculations or collision predictions
\citep[e.g.][]{Virtanen2006,Oszkiewicz2012,Farnocchia2015}.
These orbits cover a wide parameter space with
semi-major axes between 0.6 and 1000\,au, eccentricities between 0.0 and 1.0, inclinations between 0 and
180$^{\circ}$, and perihelion distances are from 0.008 to 29\,au. All of these orbits can explain the
MIRI-specific astrometric data presented in Tbl.~\ref{tbl:astrometry} within assumed uncertainties
of $\pm$0.025$^{\prime \prime}$ and $\pm$0.050$^{\prime \prime}$ respectively for the two objects,
with a reduced $\chi^2$ of the orbital fit being close to unity.

For the new asteroid, a similar procedure was applied, this time for the specific apparent motion
of 11.37$^{\prime \prime}$/hour.
The almost 10\,000 orbits cover a similar parameter space as the ones for (10920), but due 
to the different motion of the arc observed in the sky, the orbit ranging approach
constrained the orbital perihelia to values below 3.3\,au.

Each of the possible orbit solution is coming with orbital parameters $a$ (semi-major axis), $e$ (eccentricity),
$i$ (inclination), $q$ (perihelion distance), and the calculated $r_{helio}$ (helio-centric distance [au] from the Sun),
$\Delta$ (JWST-asteroid distance [au]), and $\alpha$ (phase angle in [$^{\circ}$])
for the specific JWST observing geometry and observing epoch.

\subsection{STM calculations and fit to the MIRI data}

The possible orbit solutions are now taken to calculate flux predictions at the MIRI reference
wavelengths, and by using the corresponding $r_{helio}$, $\Delta$, and $\alpha$ values. The
model calculations are done via the STM, but with the option to use different values for the beaming parameter $\eta$.
A 1-km diameter is used as a starting value and predictions are done for a range of different albedos
(p$_V$ = 0.05, 0.10, 0.15, 0.20, and 0.25). In a second step, the reference diameter is scaled up or down to
obtain the best fit in terms of $\chi^2$-minimization to the MIRI fluxes:
The $\chi^2$ calculation is done via $\sum \frac{(flx_{miri}(i) - flx_{STM}(i))^2}{{\sigma}(i)^2}$,
with $flx_{miri}$ and $\sigma$ being the individual MIRI fluxes and absolute errors
(see Table~\ref{tbl:asteroids_miriflx}), and $flx_{STM}$ the corresponding STM prediction.
This recipe produced for each orbit (and each albedo value) the best-fit size solution and the corresponding
(reduced) $\chi^2$ value.

\subsection{MBA (10920)}

\begin{figure}[h!tb]
	\resizebox{\hsize}{!}{\includegraphics{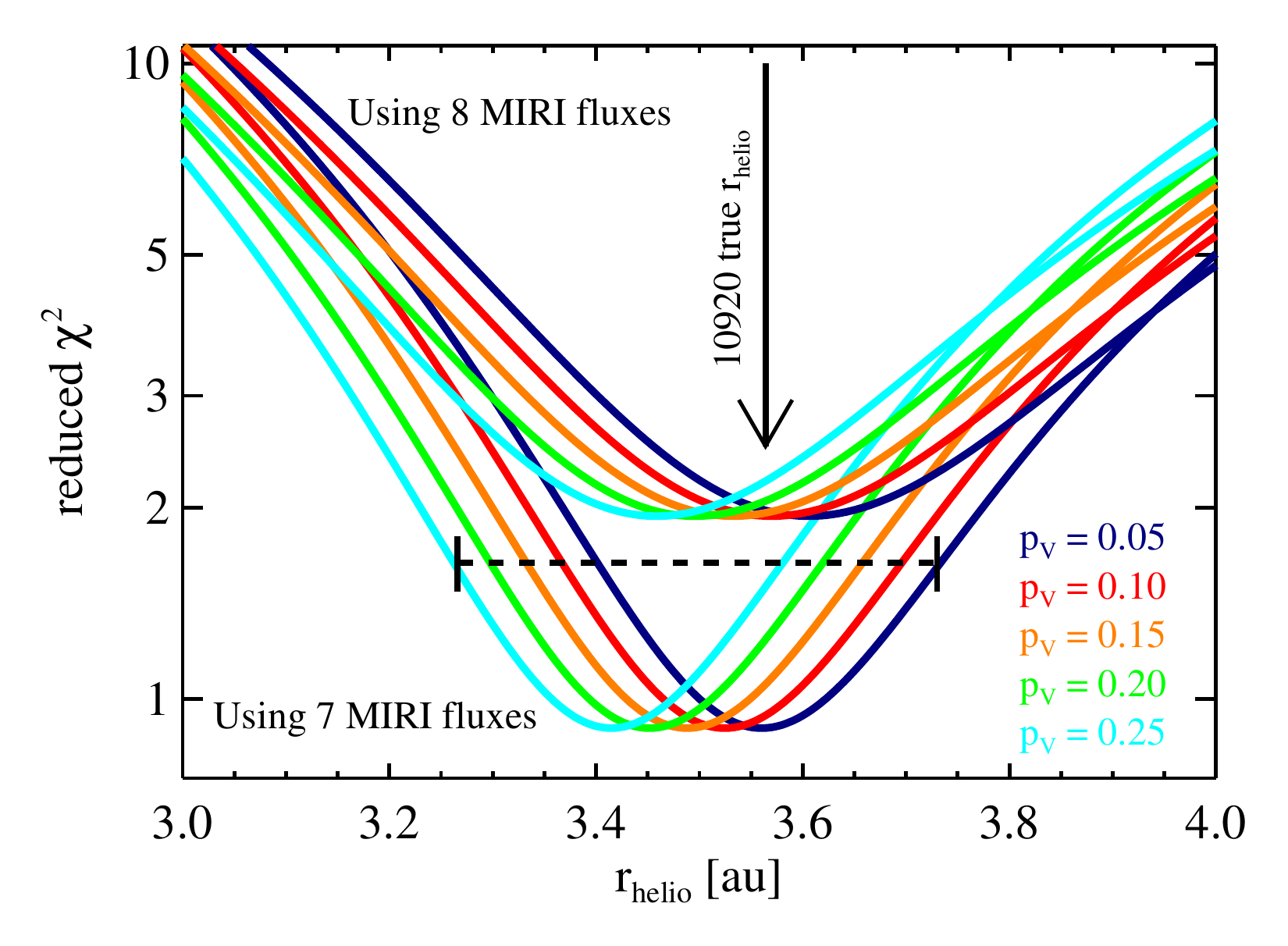}}
        \resizebox{\hsize}{!}{\includegraphics{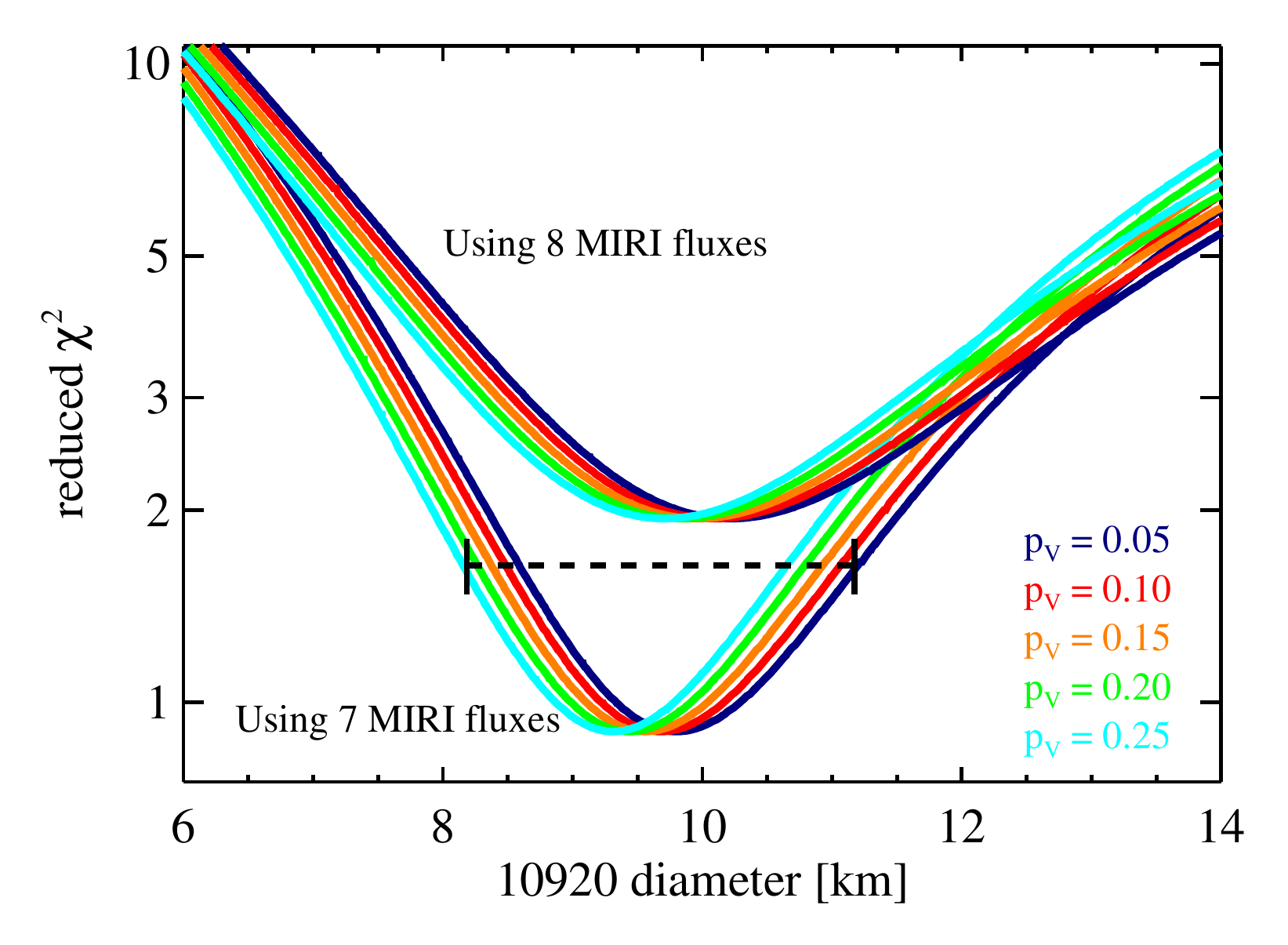}}
        \caption{Results of the $\chi^2$-fitting of STM predictions to the measured
		 MIRI fluxes of asteroid (10920). Top: Assuming that the orbit of (10920) is not known,
		 we find that the most likely r$_{helio}$ would be in the range between 3.25 and 3.75\,au.
                 Bottom: The most-likely size, based on the MIRI data alone and assuming a spherical
		 shape of (10920), would be in the range between 8 and about 11.5\,km.
         \label{fig:chi2_10920}}
\end{figure}

The $\chi^2$ fitting for (10920) was done for all 8 fluxes from 5.6 to 21.0\,$\mu$m
(see Tbl.~\ref{tbl:asteroids_miriflx}), with $\mu = n - m = 7$ as the degree of freedom
(considering the diameter scaling as the only fitting parameter $m$).
The 21-$\mu$m point is not well matched by our spherical STM shape model. The reason could 
either be the fact that (10920) was partially outside MIRI's FOV and the photometry is simply
wrong, or spin-shape related problems (see Section~\ref{sec:aux_obs}). Therefore, we did 
a separate $\chi^2$ fitting for the 7 good-quality fluxes (from 5.6 to 18.0\,$\mu$m), and
with $\mu$ = 6 as the degree of freedom for the reduced $\chi^2$ calculations. The results
are shown in Fig.~\ref{fig:chi2_10920}.

The STM-ORBIT method puts very strong constraints on the object's helio-centric distance at the
time of the JWST observations: 3.27\,au $<$ r$_{helio}$ $<$ 3.73\,au, where the lower (higher) 
boundary is connected to the assumption of a high (low) albedo of p$_V$ = 0.25 (0.05). This places
the object at the edge of the outer main-belt at the time of the JWST observations.
In a similar way, the JWST-object distance is limited to 2.56\,au $<$ $\Delta_{JWST}$ $<$ 3.04\,au,
and the phase angle $\alpha$ to values between 13.0 and 14.9$^{\circ}$. These derived radiometric
distance and angle ranges for (10920) 1998~BC1 are in excellent agreement with the true values
r$_{helio}$ = 3.56\,au, $\Delta_{JWST}$ = 2.86\,au, $\alpha$ = 13.5$^{\circ}$ at 2022-Jul-14 11:30 UT.

It is interesting to see that the small observed arc combined with the MIRI fluxes also limit the
possible orbital parameters $a$, $e$, $i$, and $q$ (see Fig.~\ref{fig:orbit_10920_chi2}):
the orbit's semi-major axis $a$ has to
be larger than 2.76\,au (true value: 3.18\,au), the eccentricity $e$ larger than 0.07 (0.15),
the inclination $i$ between 0.27$^{\circ}$ and 0.64$^{\circ}$ (0.30$^{\circ}$), and the
perihelion distance between 0.55\,au and 3.40\,au (2.70\,au). This is clearly not sufficient
for an orbit determination, but makes a classification as an outer main-belt object very likely.
It seems, we can also exclude highly eccentric orbits ($e$ $>$ 0.99) with semi-major axes
larger than about 100\,au, but here our orbit statistic was not sufficient for a solid
confirmation.

But the strongest point of the STM-ORBIT method is the size determination. Without knowing
the object's true orbit, it is possible to estimate an effective size between 8.2 and 11.2\,km,
with the smaller value being connected to p$_V$ = 0.25 and the larger one to p$_V$ = 0.05.
The STM-related size is smaller than the radiometric TPM size (see Section~\ref{sec:radiometry}),
but this is mainly related to the spherical shape (as compared to the extremely elongated
shape in the TPM study) and the fact that JWST caught (10920) during the lightcurve minimum
when its cross section is minimal.
Constraining the asteroid's albedo is not possible, as a bright and large object can produce
the same thermal emission as a dark but smaller body.

\subsection{New asteroid}

\begin{figure}[h!tb]
	\resizebox{\hsize}{!}{\includegraphics{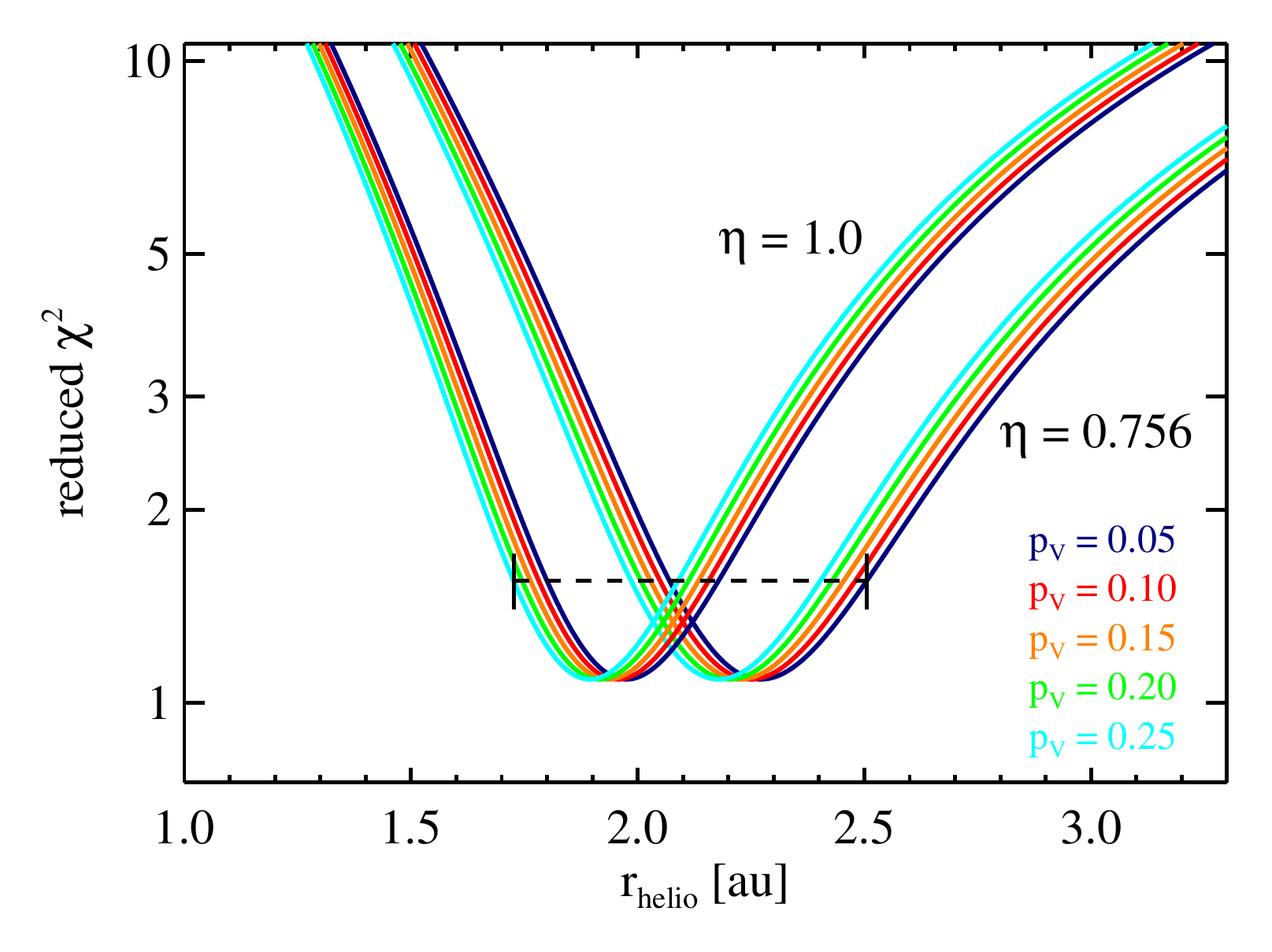}}
        \resizebox{\hsize}{!}{\includegraphics{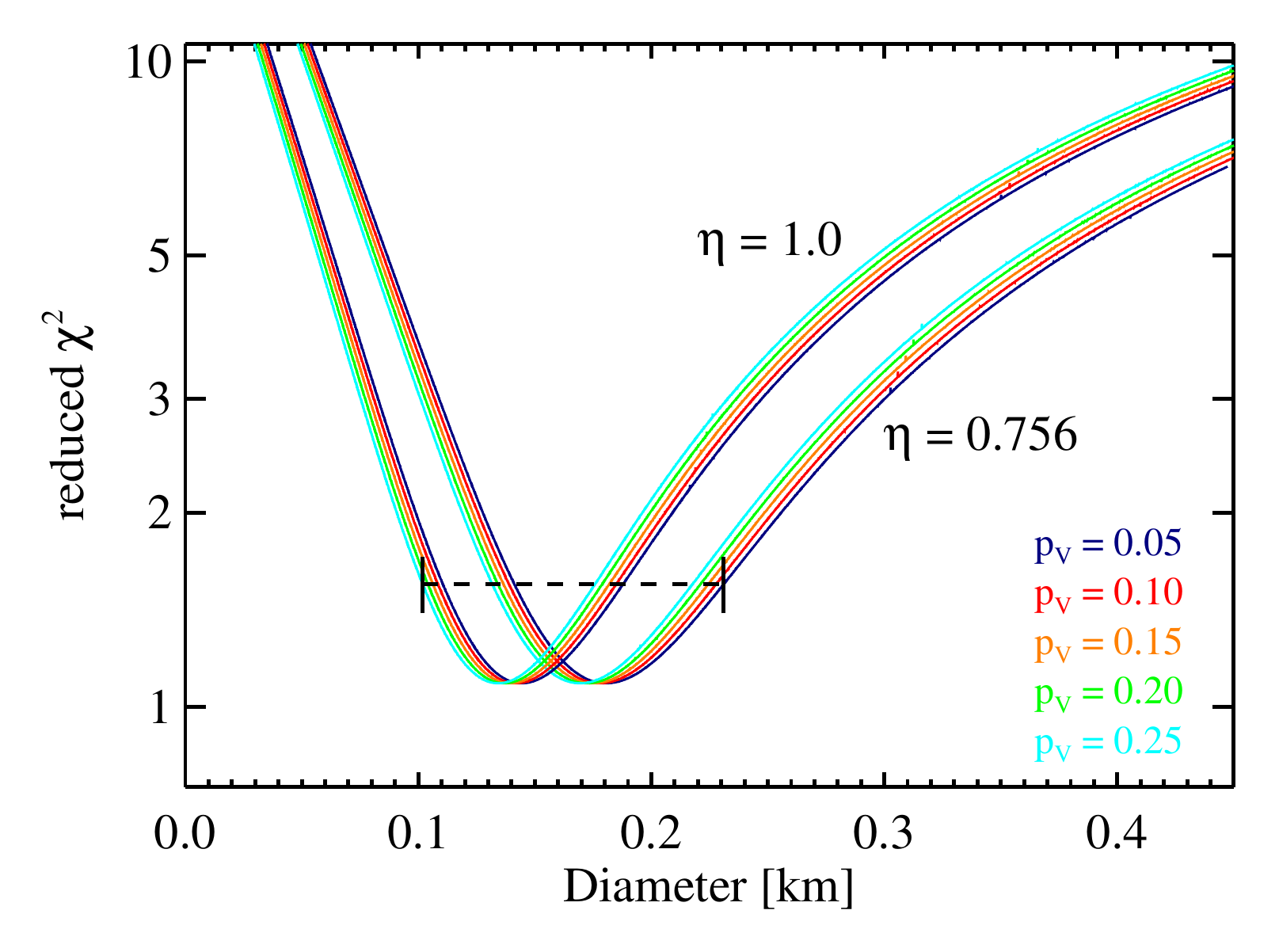}}
        \caption{Results of the $\chi^2$-fitting of STM predictions to the measured
                 MIRI fluxes of the new object. The triangles are related to a beaming
		 parameter of $\eta$ = 1.0, the diamonds belong to $\eta$ = 0.756.
		 Top: We find that the most likely r$_{helio}$ would
		 be in the range between 1.95 and 2.55\,au.
                 Bottom: The most-likely size, based on the MIRI data and assuming a spherical
		 shape, would be in the range between 100 and about 230\,m.
         \label{fig:chi2_newast}}
\end{figure}

The $\chi^2$ fitting for the new object was done for all 9 fluxes from 5.6 to 25.5\,$\mu$m
(see Tbl.~\ref{tbl:asteroids_miriflx}), with $\mu = n - m = 8$ as the degree of freedom.
The STM-ORBIT method with $\eta$ = 0.756, as it was used and verified for the outer MBA (10920), 
revealed that the new object must be located at a heliocentric distance of about 2.0 - 2.5\,au.
As smaller objects located closer to the Sun tend to have larger beaming values (see, e.g.,
discussions in \citet{Ali-Lagoa2016,Ali-Lagoa2017}), we also used $\eta$ = 1.0. It turned out
that both $\eta$-values can fit the 9 fluxes equally well, but the related best-fit helio-centric
distances differ (see Fig.~\ref{fig:chi2_newast}). For constraining the object's size, location and orbit, we therefore used
the full $\eta$-range from 0.76 to 1.0. Two extreme solutions are
shown in Fig.~\ref{fig:newast_sed}: a dark (p$_V$ = 0.05, D$_{eff}$ = 180\,m) object at large
heliocentric distance of r$_{helio}$ = 2.266\,au (as red solid line) with a model beaming
parameter $\eta$ = 0.76, and a bright object (p$_V$ = 0.25, D$_{eff}$ = 151\,m) at
r$_{helio}$ = 1.975\,au (solid blue line), but with a beaming parameter of 1.0. Both
solutions fit nicely the MIRI fluxes. All $\chi^2$-compatible solutions (here, for
reduced $\chi^2$ values of 1.55) are indicated by the green lines.

The nine MIRI fluxes can then be explained best if the new object is located at 1.73\,au $<$ r$_{helio}$ $<$ 2.50\,au,
0.91\,au $<$ $\Delta_{JWST}$ $<$ 1.76\,au, 19.2$^{\circ}$ $<$ $\alpha$ $<$ 29.0$^{\circ}$.
The orbital parameters are restricted to $a$ $>$ 1.3\,au, $e$ $>$ 0.01, 0.7$^{\circ}$ $<$ $i$ $<$ 2.0$^{\circ}$,
and 0.1\,au $<$ $q$ $<$ 2.50\,au. This gives a high probability for a low-inclination,
inner main-belt object.
The size of this newly discovered small body is between 100\,m and 230\,m in diameter where the
smallest value is connected to $\eta = 1.0$ and high albedo, the largest size to $\eta = 0.756$
and a dark albedo.
A wide range for the beaming parameters $\eta$ would broaden the 
parameter range, but due to the very-likely inner main-belt location (coming with moderate
phase angles below 30$^{\circ}$) justifies the applied beaming range. However, an even larger
beaming value cannot be excluded and the object would fall into the sub-100\,m category,
with the heliocentric distance approaching 1.5\,au. We also applied the STM-ORBIT method
to a reduced observational data set. With less and less data, the $\chi^2$ goes down and 
broader and broader parameter ranges (for r$_{helio}$, D$_{eff}$) are compatible with the
observations. E.g., without the 5.6\,$\mu$m data point the possible distance from the Sun
would slightly decrease to 1.65\,au, at the same time, the connected diameter of the new
object would shift to 80-220\,m. It is therefore essential for the success of the STM-ORBIT
method to have detections in the widest possible wavelength range available.

\section{Discussion}
\label{sec:discussion}

Our STM-ORBIT method can locate the position ($r_{helio}$, $\Delta$, $\alpha$) of an unknown
object and distinguish between a near-Earth object, inner-, middle-, outer main-belt, or an object beyond the main belt.
Depending on the length of the observed arc, also the orbit inclination can be derived,
and strong limitations of the possible perihelion distances can be given.
For $a$ and $e$ the method allows to exclude some extreme cases.
However, the big advantage is the radiometric size determination without knowing the
object's true orbit. The main uncertainty is related to the poorly constrained
beaming parameter $\eta$ and the degeneracy between the object's heliocentric
distance and the applied model beaming parameter (see Figs.~\ref{fig:newast_sed} and \ref{fig:chi2_newast}).
As a starting point, we used the STM-default value of $\eta = 0.756$ \citep{Lebofsky1986}.
This is also the best-fit value for (10920) when we use the object's true distances and angles
(see Section~\ref{sec:radiometry}).
\citet{Masiero2011} used a huge WISE database to determine the beaming parameter of
main-belt asteroids as a function of phase angle $\alpha$:
$\eta$ = (0.79 $\pm$ 0.01) + $\alpha$(0.011 $\pm$ 0.001), however, with a very large
scatter of values between 0.6 and about 2.0. The formula would
lead to an $\eta$-value close to 0.95 for (10920) which is too high for our MIRI fluxes.
The high-SNR AKARI asteroid measurements \citep{Usui2011} led to a slightly different
correlation of $\eta$ = (0.76 $\pm$ 0.03) + $\alpha$(0.009 $\pm$ 0.001) \citep{Ali-Lagoa2018}
and about 5\% lower beaming values for our case here.
\citet{Grav2012} found for the more-distant Hilda group $\eta$ = 0.85 $\pm$ 0.12. As the
beaming parameter is also influenced by the covered wavelength range (with respect to the
object's thermal emission peak), and based on the fit to the (10920) MIRI fluxes, we considered
the default STM value a solid starting point for a "blind" interpretation of MIRI data.
For the maximum of $\eta$ we used 1.0, compatible with the above mentioned published correlations
for typical small-size main-belt objects. However, if an object turns out to be closer to the Sun
and/or observed under much larger phase angles, then also larger $\eta$-values have to 
be considered for the STM-ORBIT method.

For a successful application of the STM-ORBIT method it is also mandatory that
the measurements cover a wide wavelength range, or more specific, that the 
measurements allow to estimte the object's temperature (in the STM context: the
sub-solar temperature T$_{SS}$). Already a restriction to the
first 5 filters from 5.6 to 12.8\,$\mu$m would make the temperature determination
more uncertain and enlarge the possible size 
solutions by almost a factor of 2. And, the object location could be anywhere
within the main-belt. Future projects, like the NEO Surveyor\footnote{\url{https://neos.arizona.edu}}
\citep{Mainzer2015a} plan to measure only in two bands at 4-5.2\,$\mu$m and 6-10\,$\mu$m. A
similar STM-ORBIT application for newly discovered objects would then only work when multiple,
time-separated detections in both bands are available.

The unprecedented sensitivity of MIRI will guarantee that many small asteroids
will be detected. A powerful method to find and extract moving targets from the MIRI
imaging data will be the ESASky\footnote{\url{https://sky.esa.int/esasky/}. ESASky was developed
by the ESAC Science Data Centre (ESDC) team and maintained alongside other ESA science mission's
archives at ESA's European Space Astronomy Centre (ESAC, Madrid, Spain).} tool \citep{Racero2022}.
The STM-ORBIT method will allow to determine basic properties
and put constraints on their orbits. This is needed in the context of 
asteroid size-frequency distribution \citep[see e.g.,][]{Bottke2005,Bottke2015a}.
\citet{Tedesco2002c} used the ISO satellite \citep{Kessler1996} to obtain a deep asteroid
search at 12\,$\mu$m in the ecliptic plane. They estimated a cumulative number of MBAs
with diameters larger than 1\,km between 0.7 and 1.7 million objects. However, the authors
also stated that different statistical asteroid models
differ already by a factor of two at the 1 km size limit while at 2 km sizes they
are still in good agreement. The MBA statistics at a size range well below 1 km
remains unknown.
A Spitzer \citep{Werner2004} study by \citet{Ryan2009,Ryan2015} looked into the 0.5 - 1.0 km diameter asteroid  
population at different ecliptic latitudes $\beta$ between -17$^{\circ}$ and +15$^{\circ}$.
Their IR measurements
indicated that the number densities are about a factor of 2-3 below predictions by a 
Standard Asteroid Model. However, they speculated that the limiting magnitude of 
current asteroid surveys might have caused the offset.
\citet{Masiero2011} presented sizes and albedos of more than 100\,000 MBAs derived from the 
WISE/NEOWISE surveys at thermal IR wavelengths. But with very few exceptions, the derived
sizes are above 1 km. The mean beaming parameter $\eta$ was found to be about 1.0, ranging
from $\approx$ 0.94 for small phase angles to values above one at high phases
(see also our discussion above about the selection of $\eta$ for our two targets).
They confirm a bimodal albedo distribution, and decreasing average albedos when going
from the inner main belt to outer regions. But due to the WISE detection limits,
the sub-kilometer MBA regime could not be characterized.
A recent asteroid study with the Hubble Space Telescope \citep{Kruk2022} found about
60 asteroids per square degree with magnitudes brighter than 24.5\,mag in a 30$^{\circ}$-wide
ecliptic band. Size estimates from optical data alone are more difficult, but even the
faintest ones are larger than about 200\,m.

A main-belt model \citep{Bottke2015a} predicts about 10$^{8}$ asteroids with
sizes of 100\,m and larger. If we assume that they are equally distributed over
the ecliptic plane ($\pm$15$^{\circ}$), we find that {a typical} MIRI image
(BRIGHTSKY MIRI sub-array with 56.3$^{\prime \prime}$ $\times$ 56.3$^{\prime \prime}$ FOV)
will contain on average about two asteroids when pointing towards the ecliptic zone, and even
higher numbers directly in the ecliptic plane. However, this number is probably an upper limit
as a 100\,m object in middle or outer main-belt will be fainter and more difficult to
be seen in MIRI frames. Our two objects seen in the MIRI data from July 14, 2022 seem to
match these predictions, but (10920) was the prime target and there is only one 
obvious serendipitous object. Longer integrations (longer than the 21.6 or 8.7\,s per
dither position in our case) or more sophisticated search procedures will reveal
higher numbers of objects, including smaller sizes and/or more distant ones.

\section{Conclusions}
\label{sec:conclusion}

We present JWST-MIRI fluxes and positions for the outer MBA (10920) and an unknown object
in close apparent proximity. The observations were taken in MIRI imaging mode with the
BRIGHTSKY subarray on July 14, 2022 between 10:21 and 12:23 UT.
(10920) was detected in 8 bands between 5.6 and 21\,$\mu$m and its apparent motion was
4.33$^{\prime \prime}$/h, in perfect agreement with JWST-centric orbit calculations.
The new object is visible in all 9 MIRI bands, including also the 25.5\,$\mu$m band,
and it moved with 11.37$^{\prime \prime}$/h.
We combined the MIRI fluxes for (10920) with WISE/NEOWISE observations between 2010 and 2021
and obtained new lightcurves at visible wavelengths in August and September 2022. Lightcurve
inversion techniques and a radiometric study revealed that (10920) is very elongated
(a/b $\ge$ 1.5), rotates with 4.861191\,h, with a spin-pole at
($\lambda$,$\beta$) = (178$^{\circ}$, +81$^{\circ}$). It has a size of 14.5 - 16.5\,km
(diameter of an equal-volume sphere), a geometric albedo p$_{V}$ between 0.05 and 0.10, and
a thermal inertia in the range 9 to 35 (best value 15)\,J\,m$^{-2}$s$^{-0.5}$K$^{-1}$.
Albedo and thermal inertia are in good agreement with expectations for C-complex outer
MBAs \citep{Delbo2015}.

We used the MIRI positions and fluxes to develop a new "STM-ORBIT" method which allows
to constrain an object's helio-centric distance at the time of observation and its size,
without knowing the object's true orbit. The STM-ORBIT technique was tested and validated for
(10920) and then applied to the new unkonwn object. The new object was very likely 
located in the inner main-belt region at the time of the JWST observations, and it
is on a very low inclination orbit. It has a diameter of 100-230\,m, a size range which
is very poorly characterized, but very important for size-frequency
distribution studies. From a size-frequency model by \citet{Bottke2015} and our 
experience with the above presented data, we estimate that typical MIRI images
(with a FOV of roughly 1$^{\prime}$$\times$1$^{\prime}$) will include on average
about 1-2 objects with sizes of 100\,m or larger when pointed at low ecliptic latitude.
However, the size and position determination for objects without known orbit will only
be possible via well-characterized thermal infrared spectra or spectral slopes,
preferentially with multi-band detections close to the thermal emission peak.


\begin{acknowledgements}
  TSR acknowledges funding from the NEO-MAPP project (H2020-EU-2-1-6/870377). This work was
  (partially) funded by the Spanish MICIN/AEI/10.13039/501100011033 and by "ERDF A way of
  making Europe" by the European Union through grant RTI2018-095076-B-C21, and the Institute
  of Cosmos Sciences University of Barcelona (ICCUB, Unidad de Excelencia 'Mar\'{i}a de Maeztu')
  through grant CEX2019-000918-M.
  T.M.\ would like to thank V\'{i}ctor Al\'{i}-Lagoa for extensive discussion
  on the beaming parameter $\eta$, Esa Vilenius for his support with the 
  interpretation of $\chi^2$-related statistics, and Karl Gordon for providing information
  on MIRI photometry. Andras Pal searched through the
  TESS and Kepler K2 archives for finding more lightcurves for (10920) 1998~BC1,
  but the asteroid was not covered by these surveys.
  This publication makes use of data products from the Wide-field Infrared Survey Explorer
  (WISE), which is a joint project of the University of California, Los Angeles, and the Jet
  Propulsion Laboratory/California Institute of Technology, and data products
  from the Near-Earth Object Wide-field Infrared Survey Explorer (NEOWISE),
  which is a joint project of the Jet Propulsion Laboratory/California Institute
  of Technology and the University of Arizona. WISE and NEOWISE are funded by the
  National Aeronautics and Space Administration.
  This project used public archival data from the Dark Energy Survey (DES).
  Funding for the DES Projects has been provided by the U.S. Department of Energy, the U.S.
  National Science Foundation, the Ministry of Science and Education of Spain, the Science
  and Technology FacilitiesCouncil of the United Kingdom, the Higher Education Funding Council
  for England, the National Center for Supercomputing Applications at the University of Illinois
  at Urbana-Champaign, the Kavli Institute of Cosmological Physics at the University of Chicago,
  the Center for Cosmology and Astro-Particle Physics at the Ohio State University, the Mitchell
  Institute for Fundamental Physics and Astronomy at Texas A\&M University, Financiadora de
  Estudos e Projetos, Funda{\c c}{\~a}o Carlos Chagas Filho de Amparo {\`a} Pesquisa do Estado
  do Rio de Janeiro, Conselho Nacional de Desenvolvimento Cient{\'i}fico e Tecnol{\'o}gico
  and the Minist{\'e}rio da Ci{\^e}ncia, Tecnologia e Inova{\c c}{\~a}o, the Deutsche
  Forschungsgemeinschaft, and the Collaborating Institutions in the Dark Energy Survey.
  A large fraction of this project was conducted during a personal COVID isolation and
  quarantine phase of the first author in October 2022.
  PPB acknowledges funding through the Spanish Government retraining plan 'María Zambrano 2021-2023' at the University of Alicante
  (ZAMBRANO22-04).
\end{acknowledgements}

\clearpage
\bibliographystyle{aa}  
\bibliography{AsteroidsGeneral}

\begin{thebibliography}{57}
\expandafter\ifx\csname natexlab\endcsname\relax\def\natexlab#1{#1}\fi

\bibitem[{{Al{\'{\i}}-Lagoa} \& {Delbo'}(2017)}]{Ali-Lagoa2017}
{Al{\'{\i}}-Lagoa}, V. \& {Delbo'}, M. 2017, \aap, 603, A55

\bibitem[{{Al{\'{\i}}-Lagoa} {et~al.}(2016){Al{\'{\i}}-Lagoa}, {Licandro},
  {Gil-Hutton}, {Ca{\~n}ada-Assandri}, {Delbo'}, {de Le{\'o}n}, {Campins},
  {Pinilla-Alonso}, {Kelley}, \& {Hanu{\v s}}}]{Ali-Lagoa2016}
{Al{\'{\i}}-Lagoa}, V., {Licandro}, J., {Gil-Hutton}, R., {et~al.} 2016, \aap,
  591, A14

\bibitem[{{Al{\'\i}-Lagoa} {et~al.}(2018){Al{\'\i}-Lagoa}, {M{\"u}ller},
  {Usui}, \& {Hasegawa}}]{Ali-Lagoa2018}
{Al{\'\i}-Lagoa}, V., {M{\"u}ller}, T.~G., {Usui}, F., \& {Hasegawa}, S. 2018,
  \aap, 612, A85

\bibitem[{{Bartczak} \& {Dudzi{\'n}ski}(2018)}]{Bartczak2018}
{Bartczak}, P. \& {Dudzi{\'n}ski}, G. 2018, \mnras, 473, 5050

\bibitem[{{Bottke} {et~al.}(2015{\natexlab{a}}){Bottke}, {Bro{\v{z}}},
  {O'Brien}, {Campo Bagatin}, {Morbidelli}, \& {Marchi}}]{Bottke2015a}
{Bottke}, W.~F., {Bro{\v{z}}}, M., {O'Brien}, D.~P., {et~al.}
  2015{\natexlab{a}}, in Asteroids IV, ed. P.~{Michel}, F.~{DeMeo}, \&
  W.~{Bottke} ({University of Arizona Press}), 701--724

\bibitem[{{Bottke} {et~al.}(2005){Bottke}, {Durda}, {Nesvorn{\'y}}, {Jedicke},
  {Morbidelli}, {Vokrouhlick{\'y}}, \& {Levison}}]{Bottke2005}
{Bottke}, W.~F., {Durda}, D.~D., {Nesvorn{\'y}}, D., {et~al.} 2005, \icarus,
  175, 111

\bibitem[{{Bottke} {et~al.}(2015{\natexlab{b}}){Bottke}, {Vokrouhlick{\'y}},
  {Walsh}, {Delbo}, {Michel}, {Lauretta}, {Campins}, {Connolly}, {Scheeres}, \&
  {Chelsey}}]{Bottke2015}
{Bottke}, W.~F., {Vokrouhlick{\'y}}, D., {Walsh}, K.~J., {et~al.}
  2015{\natexlab{b}}, \icarus, 247, 191

\bibitem[{{Bouchet} {et~al.}(2015){Bouchet}, {Garc{\'\i}a-Mar{\'\i}n},
  {Lagage}, {Amiaux}, {Augu{\'e}res}, {Bauwens}, {Blommaert}, {Chen}, {Detre},
  {Dicken}, {Dubreuil}, {Galdemard}, {Gastaud}, {Glasse}, {Gordon}, {Gougnaud},
  {Guillard}, {Justtanont}, {Krause}, {Leboeuf}, {Longval}, {Martin}, {Mazy},
  {Moreau}, {Olofsson}, {Ray}, {Rees}, {Renotte}, {Ressler}, {Ronayette},
  {Salasca}, {Scheithauer}, {Sykes}, {Thelen}, {Wells}, {Wright}, \&
  {Wright}}]{Bouchet2015}
{Bouchet}, P., {Garc{\'\i}a-Mar{\'\i}n}, M., {Lagage}, P.~O., {et~al.} 2015,
  \pasp, 127, 612

\bibitem[{{Cellino} {et~al.}(2015){Cellino}, {Muinonen}, {Hestroffer}, \&
  {Carbognani}}]{Cellino2015}
{Cellino}, A., {Muinonen}, K., {Hestroffer}, D., \& {Carbognani}, A. 2015,
  \planss, 118, 221

\bibitem[{{Delbo} {et~al.}(2015){Delbo}, {Mueller}, {Emery}, {Rozitis}, \&
  {Capria}}]{Delbo2015}
{Delbo}, M., {Mueller}, M., {Emery}, J.~P., {Rozitis}, B., \& {Capria}, M.~T.
  2015, {Asteroid Thermophysical Modeling} ({University of Arizona Press,
  Tucson}), 107--128

\bibitem[{{Farnocchia} {et~al.}(2015){Farnocchia}, {Chesley}, \&
  {Micheli}}]{Farnocchia2015}
{Farnocchia}, D., {Chesley}, S.~R., \& {Micheli}, M. 2015, \icarus, 258, 18

\bibitem[{{Flaugher} {et~al.}(2015){Flaugher}, {Diehl}, {Honscheid}, {Abbott},
  {Alvarez}, {Angstadt}, {Annis}, {Antonik}, {Ballester}, {Beaufore},
  {Bernstein}, {Bernstein}, {Bigelow}, {Bonati}, {Boprie}, {Brooks},
  {Buckley-Geer}, {Campa}, {Cardiel-Sas}, {Castander}, {Castilla}, {Cease},
  {Cela-Ruiz}, {Chappa}, {Chi}, {Cooper}, {da Costa}, {Dede}, {Derylo},
  {DePoy}, {de Vicente}, {Doel}, {Drlica-Wagner}, {Eiting}, {Elliott}, {Emes},
  {Estrada}, {Fausti Neto}, {Finley}, {Flores}, {Frieman}, {Gerdes},
  {Gladders}, {Gregory}, {Gutierrez}, {Hao}, {Holland}, {Holm}, {Huffman},
  {Jackson}, {James}, {Jonas}, {Karcher}, {Karliner}, {Kent}, {Kessler},
  {Kozlovsky}, {Kron}, {Kubik}, {Kuehn}, {Kuhlmann}, {Kuk}, {Lahav}, {Lathrop},
  {Lee}, {Levi}, {Lewis}, {Li}, {Mandrichenko}, {Marshall}, {Martinez},
  {Merritt}, {Miquel}, {Mu{\~n}oz}, {Neilsen}, {Nichol}, {Nord}, {Ogando},
  {Olsen}, {Palaio}, {Patton}, {Peoples}, {Plazas}, {Rauch}, {Reil}, {Rheault},
  {Roe}, {Rogers}, {Roodman}, {Sanchez}, {Scarpine}, {Schindler}, {Schmidt},
  {Schmitt}, {Schubnell}, {Schultz}, {Schurter}, {Scott}, {Serrano}, {Shaw},
  {Smith}, {Soares-Santos}, {Stefanik}, {Stuermer}, {Suchyta}, {Sypniewski},
  {Tarle}, {Thaler}, {Tighe}, {Tran}, {Tucker}, {Walker}, {Wang}, {Watson},
  {Weaverdyck}, {Wester}, {Woods}, {Yanny}, \& {DES
  Collaboration}}]{Flaugher2015}
{Flaugher}, B., {Diehl}, H.~T., {Honscheid}, K., {et~al.} 2015, \aj, 150, 150

\bibitem[{{Ginsburg} {et~al.}(2019){Ginsburg}, {Sip{\H{o}}cz}, {Brasseur},
  {Cowperthwaite}, {Craig}, {Deil}, {Guillochon}, {Guzman}, {Liedtke}, {Lian
  Lim}, {Lockhart}, {Mommert}, {Morris}, {Norman}, {Parikh}, {Persson},
  {Robitaille}, {Segovia}, {Singer}, {Tollerud}, {de Val-Borro}, {Valtchanov},
  {Woillez}, {Astroquery Collaboration}, \& {a subset of astropy
  Collaboration}}]{Ginsburg2019}
{Ginsburg}, A., {Sip{\H{o}}cz}, B.~M., {Brasseur}, C.~E., {et~al.} 2019, \aj,
  157, 98

\bibitem[{{Gordon} {et~al.}(2022){Gordon}, {Bohlin}, {Sloan}, {Rieke}, {Volk},
  {Boyer}, {Muzerolle}, {Schlawin}, {Deustua}, {Hines}, {Kraemer}, {Mullally},
  \& {Su}}]{Gordon2022}
{Gordon}, K.~D., {Bohlin}, R., {Sloan}, G.~C., {et~al.} 2022, \aj, 163, 267

\bibitem[{{Grav} {et~al.}(2012){Grav}, {Mainzer}, {Bauer}, {Masiero}, {Spahr},
  {McMillan}, {Walker}, {Cutri}, {Wright}, {Eisenhardt}, {Blauvelt}, {DeBaun},
  {Elsbury}, {Gautier}, {Gomillion}, {Hand}, \& {Wilkins}}]{Grav2012}
{Grav}, T., {Mainzer}, A.~K., {Bauer}, J., {et~al.} 2012, \apj, 744, 197

\bibitem[{{Harris}(1998)}]{Harris1998}
{Harris}, A.~W. 1998, \icarus, 131, 291

\bibitem[{{Kessler} {et~al.}(1996){Kessler}, {Steinz}, {Anderegg}, {Clavel},
  {Drechsel}, {Estaria}, {Faelker}, {Riedinger}, {Robson}, {Taylor}, \&
  {Xim{\'e}nez de Ferr{\'a}n}}]{Kessler1996}
{Kessler}, M.~F., {Steinz}, J.~A., {Anderegg}, M.~E., {et~al.} 1996, \aap, 315,
  L27

\bibitem[{{Kruk} {et~al.}(2022){Kruk}, {Garc{\'\i}a Mart{\'\i}n}, {Popescu},
  {Mer{\'\i}n}, {Mahlke}, {Carry}, {Thomson}, {Karada{\u{g}}}, {Dur{\'a}n},
  {Racero}, {Giordano}, {Baines}, {de Marchi}, \& {Laureijs}}]{Kruk2022}
{Kruk}, S., {Garc{\'\i}a Mart{\'\i}n}, P., {Popescu}, M., {et~al.} 2022, \aap,
  661, A85

\bibitem[{{Lagerros}(1996)}]{Lagerros1996}
{Lagerros}, J.~S.~V. 1996, \aap, 310, 1011

\bibitem[{{Lagerros}(1997)}]{Lagerros1997}
{Lagerros}, J.~S.~V. 1997, \aap, 325, 1226

\bibitem[{{Lagerros}(1998)}]{Lagerros1998}
{Lagerros}, J.~S.~V. 1998, \aap, 332, 1123

\bibitem[{{Lebofsky} {et~al.}(1986){Lebofsky}, {Sykes}, {Tedesco}, {Veeder},
  {Matson}, {Brown}, {Gradie}, {Feierberg}, \& {Rudy}}]{Lebofsky1986}
{Lebofsky}, L.~A., {Sykes}, M.~V., {Tedesco}, E.~F., {et~al.} 1986, \icarus,
  68, 239

\bibitem[{{Lu} \& {Jewitt}(2019)}]{Lu2019}
{Lu}, X.-P. \& {Jewitt}, D. 2019, \aj, 158, 220

\bibitem[{{Mahlke} {et~al.}(2021){Mahlke}, {Carry}, \& {Denneau}}]{Mahlke2021}
{Mahlke}, M., {Carry}, B., \& {Denneau}, L. 2021, \icarus, 354, 114094

\bibitem[{{Mainzer} {et~al.}(2014){Mainzer}, {Bauer}, {Cutri}, {Dailey},
  {Grav}, {Masiero}, {Nugent}, {Stevenson}, {Sonnett}, \&
  {Wright}}]{Mainzer2014}
{Mainzer}, A., {Bauer}, J., {Cutri}, R., {et~al.} 2014, in Lunar and Planetary
  Science Conference, Vol.~45, Lunar and Planetary Science Conference, 2724

\bibitem[{{Mainzer} {et~al.}(2015){Mainzer}, {Grav}, {Bauer}, {Conrow},
  {Cutri}, {Dailey}, {Fowler}, {Giorgini}, {Jarrett}, {Masiero}, {Spahr},
  {Statler}, \& {Wright}}]{Mainzer2015a}
{Mainzer}, A., {Grav}, T., {Bauer}, J., {et~al.} 2015, \aj, 149, 172

\bibitem[{{Mainzer} {et~al.}(2011){Mainzer}, {Grav}, {Masiero}, {Bauer},
  {Wright}, {Cutri}, {McMillan}, {Cohen}, {Ressler}, \&
  {Eisenhardt}}]{Mainzer2011ApJ736}
{Mainzer}, A., {Grav}, T., {Masiero}, J., {et~al.} 2011, \apj, 736, 100

\bibitem[{{Mainzer} {et~al.}(2019){Mainzer}, {Bauer}, {Cutri}, {Grav},
  {Kramer}, {Masiero}, {Sonnett}, \& {Wright}}]{Mainzer2019}
{Mainzer}, A.~K., {Bauer}, J.~M., {Cutri}, R.~M., {et~al.} 2019, NASA Planetary
  Data System

\bibitem[{{Masiero} {et~al.}(2011){Masiero}, {Mainzer}, {Grav}, {Bauer},
  {Cutri}, {Dailey}, {Eisenhardt}, {McMillan}, {Spahr}, {Skrutskie}, {Tholen},
  {Walker}, {Wright}, {DeBaun}, {Elsbury}, {Gautier}, {Gomillion}, \&
  {Wilkins}}]{Masiero2011}
{Masiero}, J.~R., {Mainzer}, A.~K., {Grav}, T., {et~al.} 2011, \apj, 741, 68

\bibitem[{{Muinonen} {et~al.}(2010){Muinonen}, {Belskaya}, {Cellino},
  {Delb{\`o}}, {Levasseur-Regourd}, {Penttil{\"a}}, \&
  {Tedesco}}]{Muinonen2010}
{Muinonen}, K., {Belskaya}, I.~N., {Cellino}, A., {et~al.} 2010, \icarus, 209,
  542

\bibitem[{{M{\"u}ller}(2002)}]{Mueller2002}
{M{\"u}ller}, T.~G. 2002, Meteoritics and Planetary Science, 37, 1919

\bibitem[{{M{\"u}ller} {et~al.}(2017){M{\"u}ller}, {{\v D}urech}, {Ishiguro},
  {Mueller}, {Kr{\"u}hler}, {Yang}, {Kim}, {O'Rourke}, {Usui}, {Kiss},
  {Altieri}, {Carry}, {Choi}, {Delbo}, {Emery}, {Greiner}, {Hasegawa}, {Hora},
  {Knust}, {Kuroda}, {Osip}, {Rau}, {Rivkin}, {Schady}, {Thomas-Osip},
  {Trilling}, {Urakawa}, {Vilenius}, {Weissman}, \& {Zeidler}}]{Mueller2017}
{M{\"u}ller}, T.~G., {{\v D}urech}, J., {Ishiguro}, M., {et~al.} 2017, \aap,
  599, A103

\bibitem[{{Norwood} {et~al.}(2016){Norwood}, {Hammel}, {Milam}, {Stansberry},
  {Lunine}, {Chanover}, {Hines}, {Sonneborn}, {Tiscareno}, {Brown}, \&
  {Ferruit}}]{Norwood2016}
{Norwood}, J., {Hammel}, H., {Milam}, S., {et~al.} 2016, \pasp, 128, 025004

\bibitem[{{Nugent} {et~al.}(2015){Nugent}, {Mainzer}, {Masiero}, {Bauer},
  {Cutri}, {Grav}, {Kramer}, {Sonnett}, {Stevenson}, \& {Wright}}]{Nugent2015}
{Nugent}, C.~R., {Mainzer}, A., {Masiero}, J., {et~al.} 2015, \apj, 814, 117

\bibitem[{{Oszkiewicz} {et~al.}(2009){Oszkiewicz}, {Muinonen}, {Virtanen}, \&
  {Granvik}}]{Oszkiewicz2009}
{Oszkiewicz}, D., {Muinonen}, K., {Virtanen}, J., \& {Granvik}, M. 2009,
  Meteoritics and Planetary Science, 44, 1897

\bibitem[{{Oszkiewicz} {et~al.}(2012){Oszkiewicz}, {Muinonen}, {Virtanen},
  {Granvik}, \& {Bowell}}]{Oszkiewicz2012}
{Oszkiewicz}, D., {Muinonen}, K., {Virtanen}, J., {Granvik}, M., \& {Bowell},
  E. 2012, \planss, 73, 30

\bibitem[{{Oszkiewicz} {et~al.}(2011){Oszkiewicz}, {Muinonen}, {Bowell},
  {Trilling}, {Penttil{\"a}}, {Pieniluoma}, {Wasserman}, \&
  {Enga}}]{Oszkiewicz2011}
{Oszkiewicz}, D.~A., {Muinonen}, K., {Bowell}, E., {et~al.} 2011, \jqsrt, 112,
  1919

\bibitem[{{Penttil{\"a}} {et~al.}(2016){Penttil{\"a}}, {Shevchenko}, {Wilkman},
  \& {Muinonen}}]{Penttilae2016}
{Penttil{\"a}}, A., {Shevchenko}, V.~G., {Wilkman}, O., \& {Muinonen}, K. 2016,
  \planss, 123, 117

\bibitem[{{Racero} {et~al.}(2022){Racero}, {Giordano}, {Carry}, {Berthier},
  {M{\"u}ller}, {Mahlke}, {Valtchanov}, {Baines}, {Kruk}, {Mer{\'\i}n},
  {Besse}, {K{\"u}ppers}, {Puga}, {Gonz{\'a}lez N{\'u}{\~n}ez},
  {Rodr{\'\i}guez}, {de la Calle}, {L{\'o}pez-Marti}, {Norman},
  {W{\r{a}}ngblad}, {L{\'o}pez-Caniego}, \& {{\'A}lvarez Crespo}}]{Racero2022}
{Racero}, E., {Giordano}, F., {Carry}, B., {et~al.} 2022, \aap, 659, A38

\bibitem[{{Rieke} {et~al.}(2022){Rieke}, {Su}, {Sloan}, \&
  {Schlawin}}]{Rieke2022}
{Rieke}, G.~H., {Su}, K., {Sloan}, G.~C., \& {Schlawin}, E. 2022, \aj, 163, 45

\bibitem[{{Rivkin} {et~al.}(2016){Rivkin}, {Marchis}, {Stansberry}, {Takir},
  {Thomas}, \& {JWST Asteroids Focus Group}}]{Rivkin2016}
{Rivkin}, A.~S., {Marchis}, F., {Stansberry}, J.~A., {et~al.} 2016, \pasp, 128,
  018003

\bibitem[{{Rozitis} \& {Green}(2011)}]{Rozitis2011}
{Rozitis}, B. \& {Green}, S.~F. 2011, \mnras, 415, 2042

\bibitem[{{Ryan} {et~al.}(2015){Ryan}, {Mizuno}, {Shenoy}, {Woodward}, {Carey},
  {Noriega-Crespo}, {Kraemer}, \& {Price}}]{Ryan2015}
{Ryan}, E.~L., {Mizuno}, D.~R., {Shenoy}, S.~S., {et~al.} 2015, \aap, 578, A42

\bibitem[{{Ryan} {et~al.}(2009){Ryan}, {Woodward}, {Dipaolo}, {Farinato},
  {Giallongo}, {Gredel}, {Hill}, {Pedichini}, {Pogge}, \&
  {Ragazzoni}}]{Ryan2009}
{Ryan}, E.~L., {Woodward}, C.~E., {Dipaolo}, A., {et~al.} 2009, \aj, 137, 5134

\bibitem[{{Santana-Ros} {et~al.}(2015){Santana-Ros}, {Bartczak},
  {Micha{\l}owski}, {Tanga}, \& {Cellino}}]{Santana-Ros2015}
{Santana-Ros}, T., {Bartczak}, P., {Micha{\l}owski}, T., {Tanga}, P., \&
  {Cellino}, A. 2015, \mnras, 450, 333

\bibitem[{{Shevchenko} {et~al.}(2022){Shevchenko}, {Belskaya}, {Slyusarev},
  {Mikhalchenko}, {Krugly}, {Chiorny}, {Lupishko}, {Oszkiewicz}, {Kwiatkowski},
  {Gritsevich}, {Muinonen}, \& {Penttil{\"a}}}]{Shevchenko2022}
{Shevchenko}, V.~G., {Belskaya}, I.~N., {Slyusarev}, I.~G., {et~al.} 2022,
  \aap, 666, A190

\bibitem[{{Tanga} {et~al.}(2022){Tanga}, {Pauwels}, {Mignard}, {Muinonen},
  {Cellino}, {David}, {Hestroffer}, {Spoto}, {Berthier}, {Guiraud}, {Roux},
  {Carry}, {Delbo}, {Dell Oro}, {Fouron}, {Galluccio}, {Jonckheere}, {Klioner},
  {Lefustec}, {Liberato}, {Ord{\'e}novic}, {Oreshina-Slezak}, {Penttil{\"a}},
  {Pailler}, {Panem}, {Petit}, {Portell}, {Poujoulet}, {Thuillot}, {Van
  Hemelryck}, {Burlacu}, {Lasne}, \& {Managa}}]{Tanga2022}
{Tanga}, P., {Pauwels}, T., {Mignard}, F., {et~al.} 2022, arXiv e-prints,
  arXiv:2206.05561

\bibitem[{{Tedesco} \& {Desert}(2002)}]{Tedesco2002c}
{Tedesco}, E.~F. \& {Desert}, F.-X. 2002, \aj, 123, 2070

\bibitem[{{Tedesco} {et~al.}(2002{\natexlab{a}}){Tedesco}, {Egan}, \&
  {Price}}]{Tedesco2002b}
{Tedesco}, E.~F., {Egan}, M.~P., \& {Price}, S.~D. 2002{\natexlab{a}}, \aj,
  124, 583

\bibitem[{{Tedesco} {et~al.}(2002{\natexlab{b}}){Tedesco}, {Noah}, {Noah}, \&
  {Price}}]{Tedesco2002}
{Tedesco}, E.~F., {Noah}, P.~V., {Noah}, M., \& {Price}, S.~D.
  2002{\natexlab{b}}, \aj, 123, 1056

\bibitem[{{Thomas} {et~al.}(2016){Thomas}, {Abell}, {Castillo-Rogez},
  {Moskovitz}, {Mueller}, {Reddy}, {Rivkin}, {Ryan}, \&
  {Stansberry}}]{Thomas2016}
{Thomas}, C.~A., {Abell}, P., {Castillo-Rogez}, J., {et~al.} 2016, \pasp, 128,
  018002

\bibitem[{{Tonry} {et~al.}(2018){Tonry}, {Denneau}, {Heinze}, {Stalder},
  {Smith}, {Smartt}, {Stubbs}, {Weiland}, \& {Rest}}]{Tonry2018}
{Tonry}, J.~L., {Denneau}, L., {Heinze}, A.~N., {et~al.} 2018, \pasp, 130,
  064505

\bibitem[{{Usui} {et~al.}(2011){Usui}, {Kuroda}, {M{\"u}ller}, {Hasegawa},
  {Ishiguro}, {Ootsubo}, {Ishihara}, {Kataza}, {Takita}, {Oyabu}, {Ueno},
  {Matsuhara}, \& {Onaka}}]{Usui2011}
{Usui}, F., {Kuroda}, D., {M{\"u}ller}, T.~G., {et~al.} 2011, \pasj, 63, 1117

\bibitem[{{Virtanen} \& {Muinonen}(2006)}]{Virtanen2006}
{Virtanen}, J. \& {Muinonen}, K. 2006, \icarus, 184, 289

\bibitem[{{Virtanen} {et~al.}(2001){Virtanen}, {Muinonen}, \&
  {Bowell}}]{Virtanen2001}
{Virtanen}, J., {Muinonen}, K., \& {Bowell}, E. 2001, \icarus, 154, 412

\bibitem[{{Werner} {et~al.}(2004){Werner}, {Roellig}, {Low}, {Rieke}, {Rieke},
  {Hoffmann}, {Young}, {Houck}, {Brandl}, {Fazio}, {Hora}, {Gehrz}, {Helou},
  {Soifer}, {Stauffer}, {Keene}, {Eisenhardt}, {Gallagher}, {Gautier}, {Irace},
  {Lawrence}, {Simmons}, {Van Cleve}, {Jura}, {Wright}, \&
  {Cruikshank}}]{Werner2004}
{Werner}, M.~W., {Roellig}, T.~L., {Low}, F.~J., {et~al.} 2004, \apjs, 154, 1

\bibitem[{{Wright} {et~al.}(2010){Wright}, {Eisenhardt}, {Mainzer}, {Ressler},
  {Cutri}, {Jarrett}, {Kirkpatrick}, {Padgett}, {McMillan}, {Skrutskie},
  {Stanford}, {Cohen}, {Walker}, {Mather}, {Leisawitz}, {Gautier}, {McLean},
  {Benford}, {Lonsdale}, {Blain}, {Mendez}, {Irace}, {Duval}, {Liu}, {Royer},
  {Heinrichsen}, {Howard}, {Shannon}, {Kendall}, {Walsh}, {Larsen}, {Cardon},
  {Schick}, {Schwalm}, {Abid}, {Fabinsky}, {Naes}, \& {Tsai}}]{Wright2010}
{Wright}, E.~L., {Eisenhardt}, P.~R.~M., {Mainzer}, A.~K., {et~al.} 2010, \aj,
  140, 1868

\end{thebibliography}

\newpage
\begin{appendix}
\begin{onecolumn}

\section{Orbital constraints from MIRI data}

Here, we show the reduced $\chi^2$ values for all possible orbits as
a function of the object's semi-major axis $a$, the eccentricity $e$, 
the inclination $i$, and as a function of the perihelion distance $q$.
The orbital constraints are discussed in the main text. The colors are
representing different albedo values, as in Figs.~\ref{fig:chi2_10920} and
\ref{fig:chi2_newast}. Each individual point represents the reduced $\chi^2$-value
obtained from the comparison between the MIRI fluxes and the STM-ORBIT-based
flux predictions for a given orbit-albedo combination.

\subsection{MBA (10920) 1998~BC1}
\begin{figure}[h!tb]
        \resizebox{9.0cm}{!}{\includegraphics{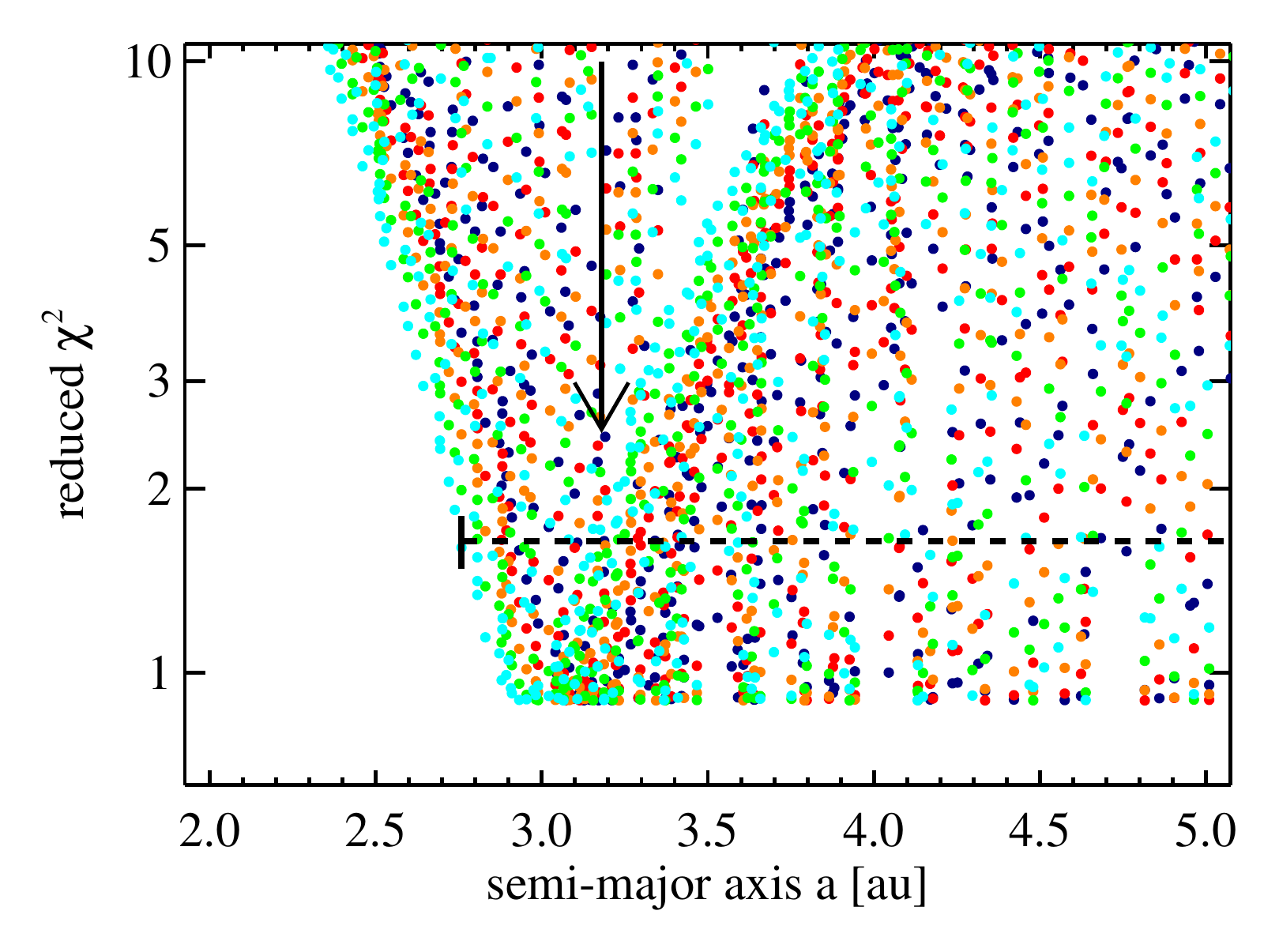}}
        \resizebox{9.0cm}{!}{\includegraphics{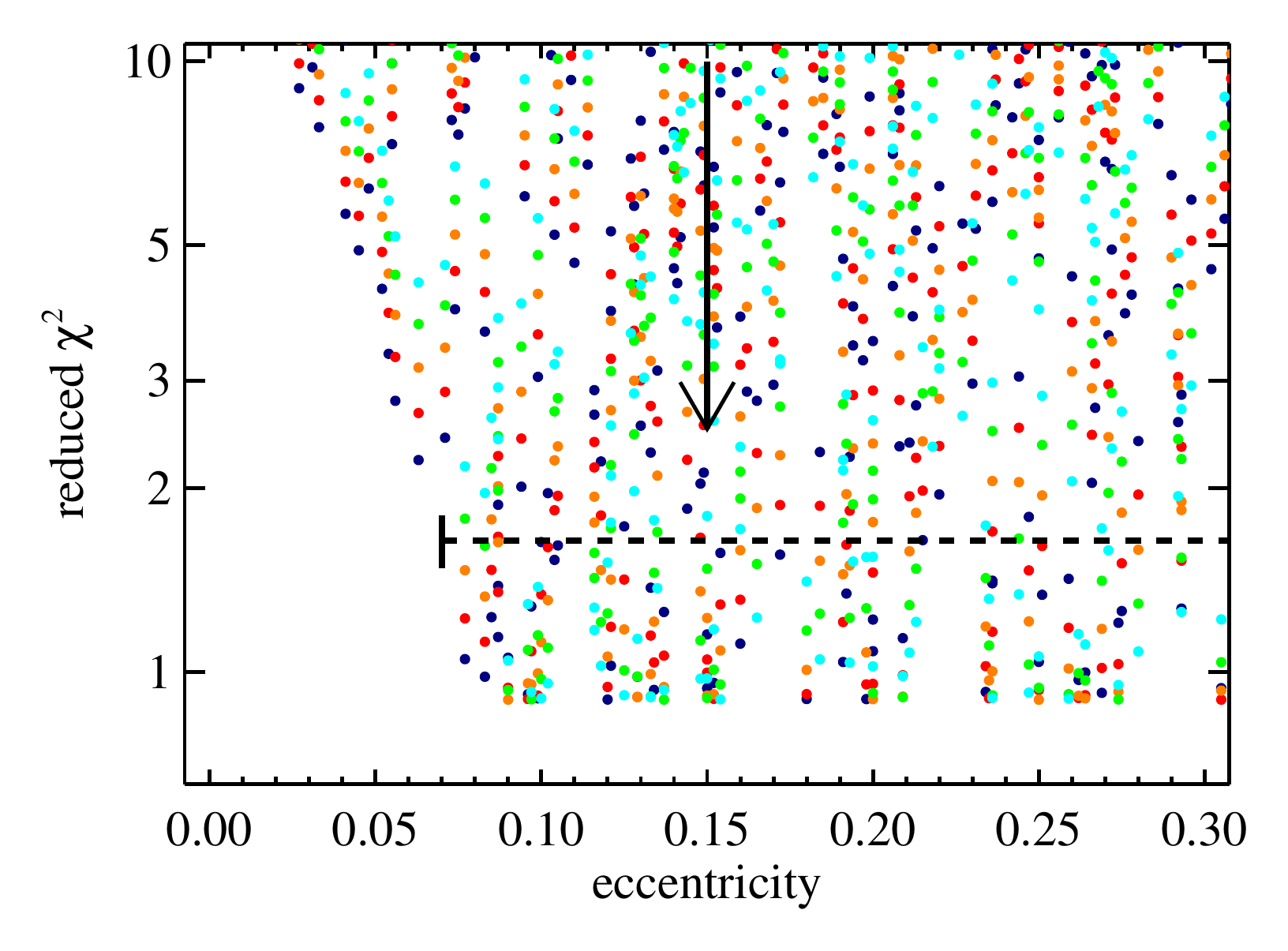}}
        \resizebox{9.0cm}{!}{\includegraphics{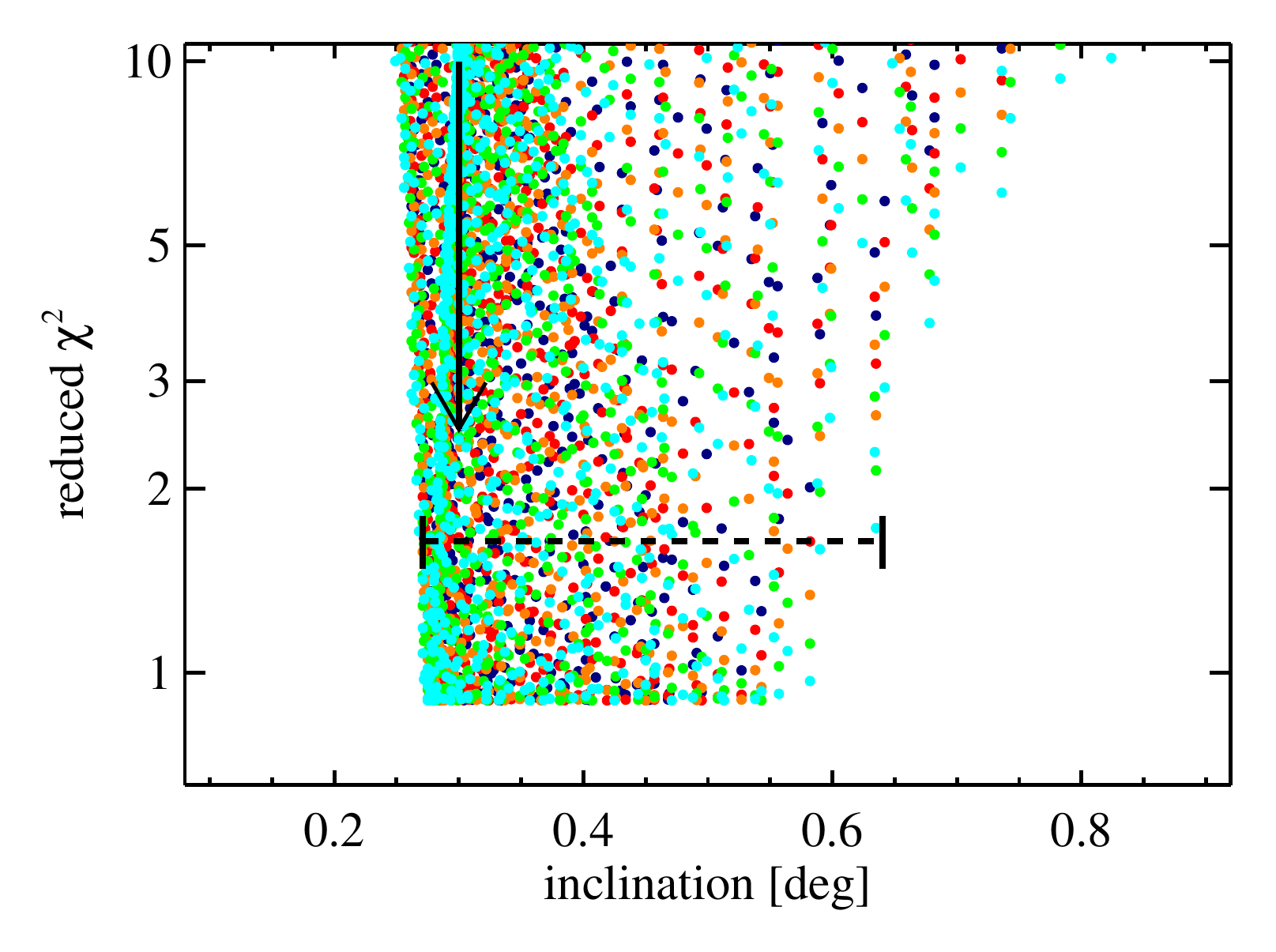}}
        \resizebox{9.0cm}{!}{\includegraphics{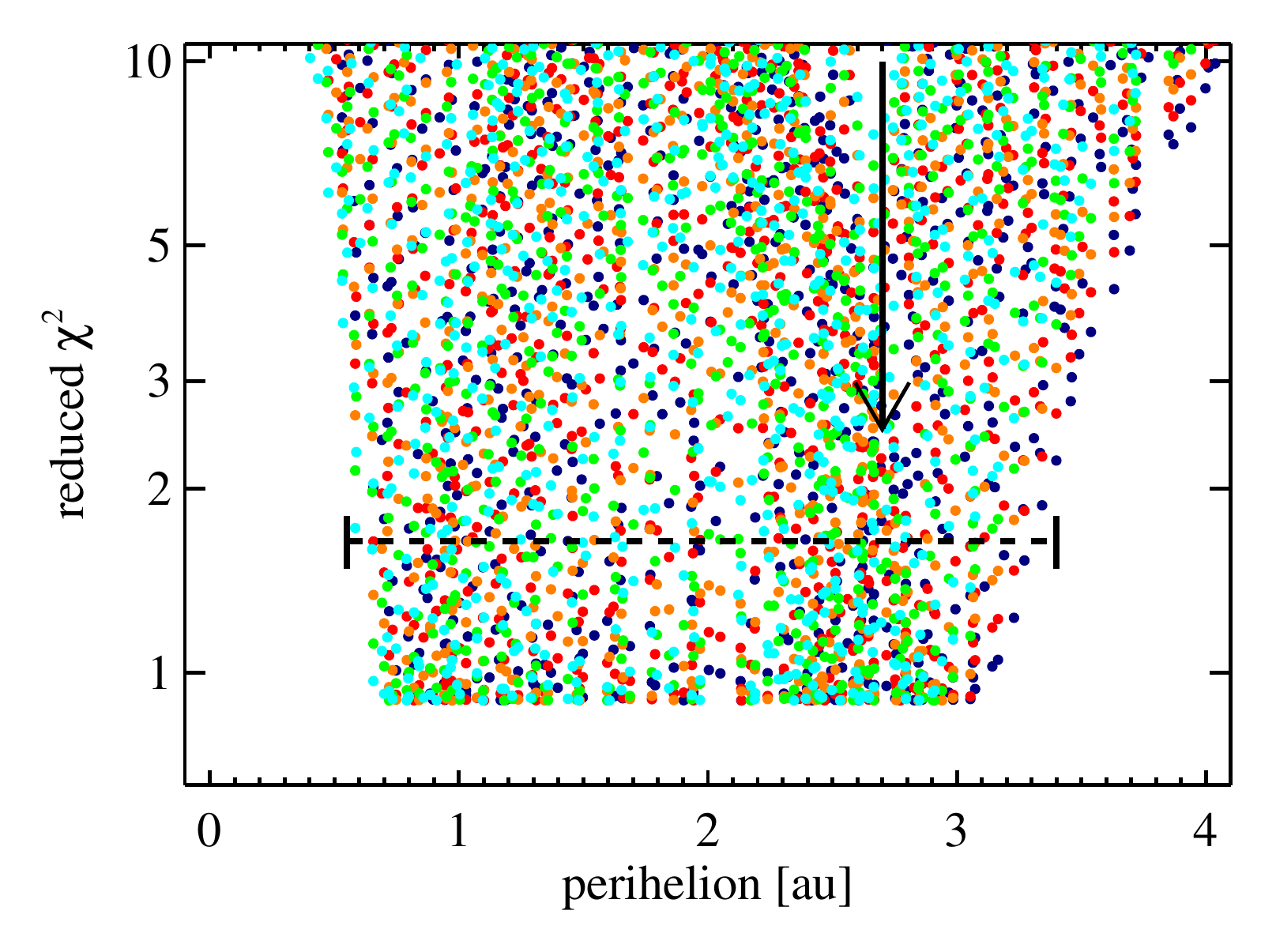}}
        \caption{Orbit constraints from the $\chi^2$-fitting of STM predictions to the measured
		 MIRI fluxes of asteroid (10920). From top to bottom: (i) Semi-major axis $a$; (ii)
                 eccentricity; (iii) inclination; (iv) perihelion distance. The true values
		 for (10920)'s orbit are indicated by a vertical arrow. The possible ranges are
                 shown at reduced $\chi^2 = 1.64$ (6 degrees of freedom).
         \label{fig:orbit_10920_chi2}}
\end{figure}

\subsection{New object}
\begin{figure}[h!tb]
	\resizebox{9.0cm}{!}{\includegraphics{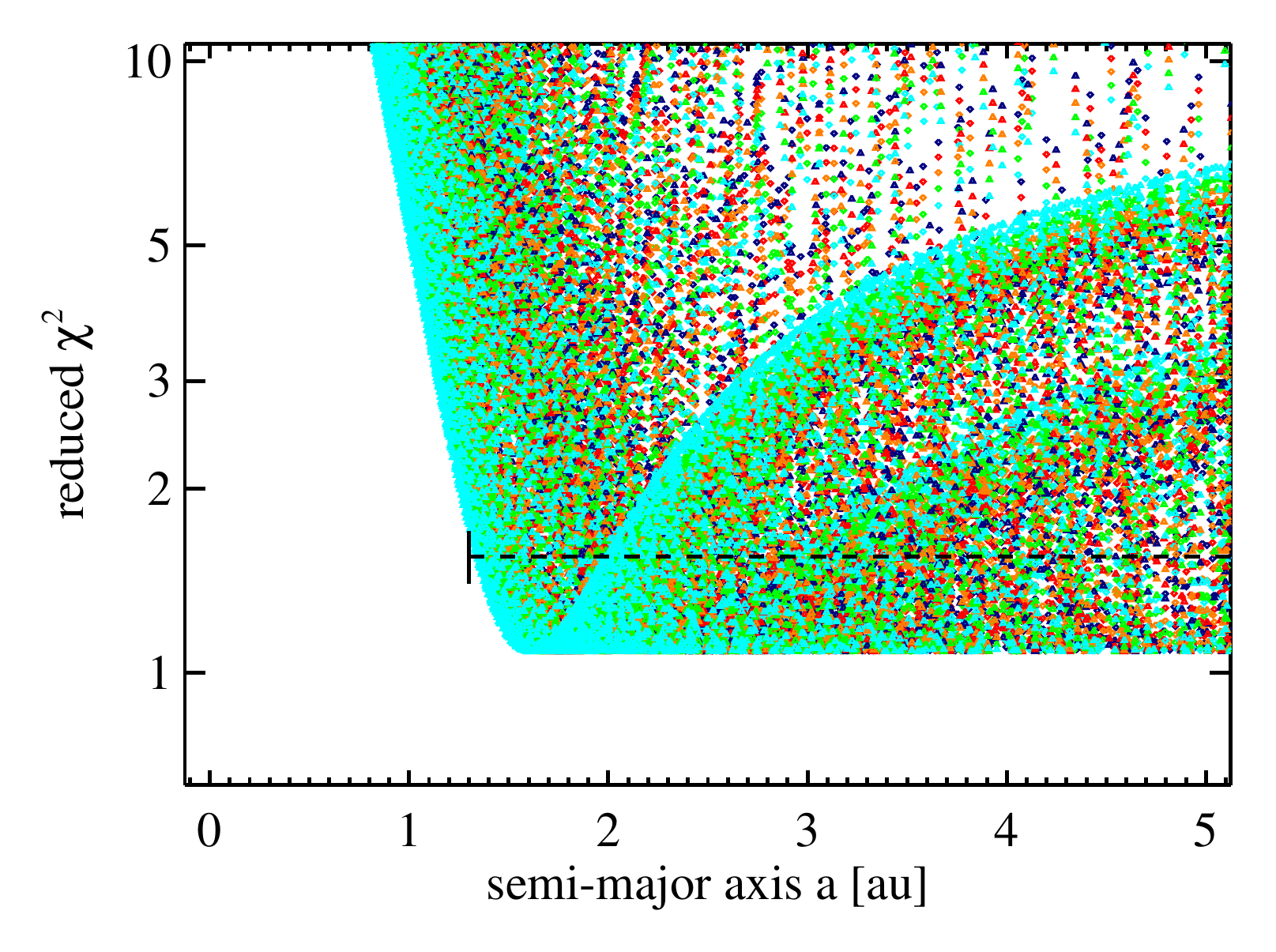}}
        \resizebox{9.0cm}{!}{\includegraphics{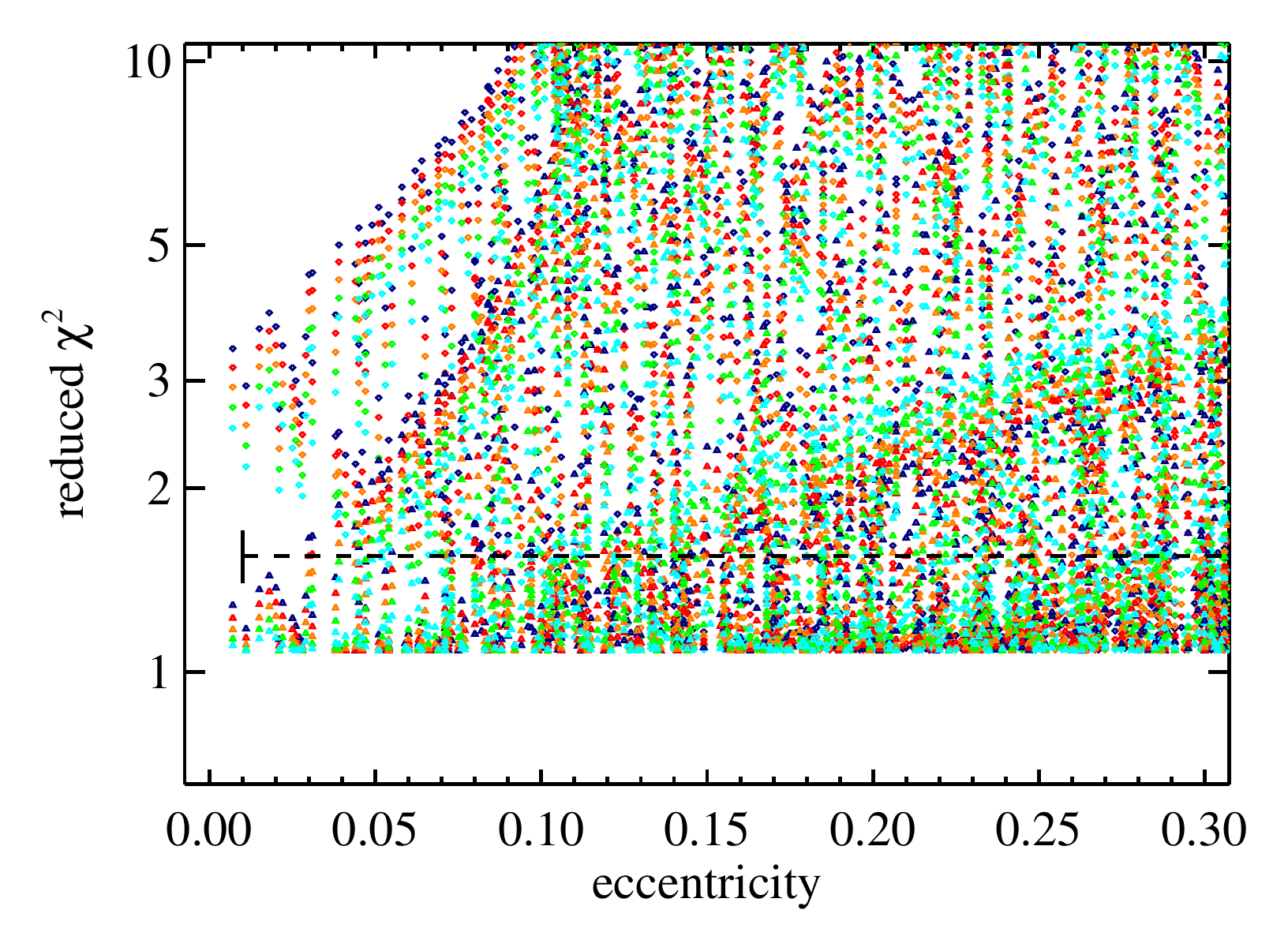}}
        \resizebox{9.0cm}{!}{\includegraphics{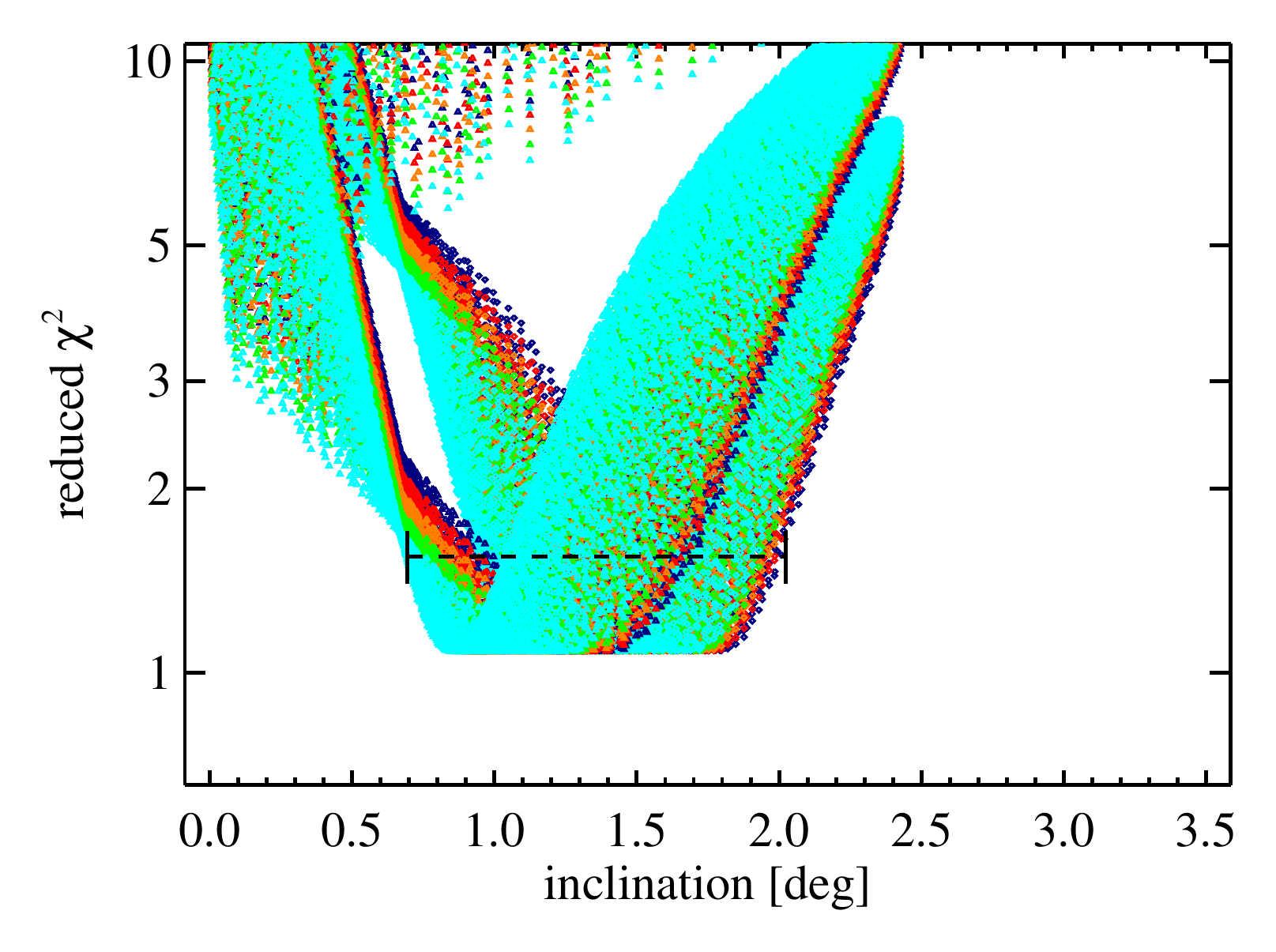}}
        \resizebox{9.0cm}{!}{\includegraphics{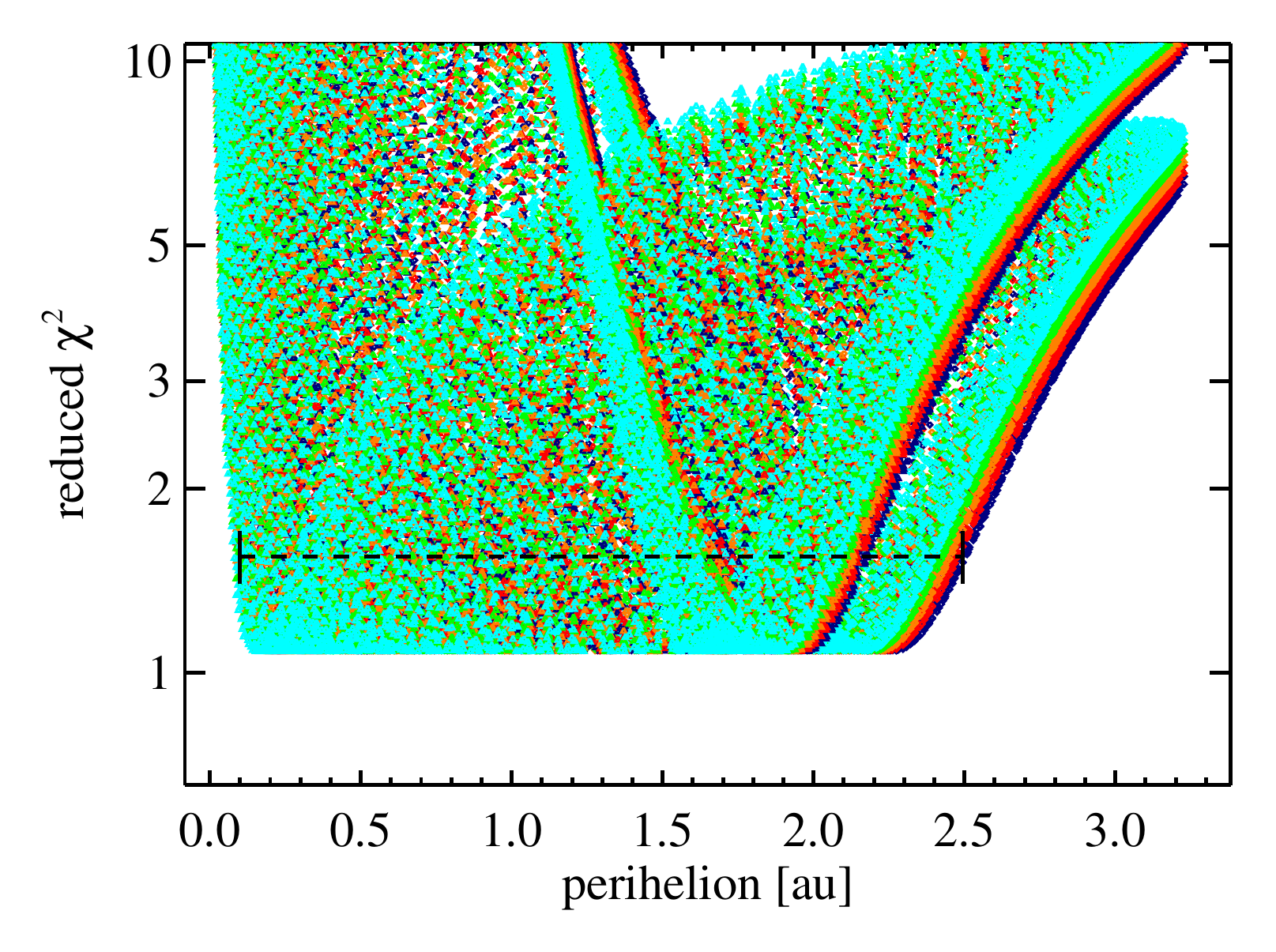}}
        \caption{Orbit constraints from the $\chi^2$-fitting of STM predictions to the measured
                 MIRI fluxes of the new object. From top to bottom: (i) Semi-major axis $a$; (ii)
                 eccentricity; (iii) inclination; (iv) perihelion distance. 
                 The possible ranges are shown at reduced $\chi^2 = 1.55$ (8 degrees of freedom).
         \label{fig:orbit_newast_chi2}}
\end{figure}

\clearpage

\section{WISE observations of MBA (10920) 1998~BC1}
\label{app:wise_obs}

We extracted all available WISE measurements for (10920) from 
the different cryogenic and post-cryogenic data sets. The magnitudes
were translated into fluxes and color-corrected to obtain monochromatic
flux densities at the WISE reference wavelengths. The error calculation
is described above. We excluded the WISE W1 band as it is dominated
by reflected sunlight for this outer main-belt object.

\begin{longtable}{lrrrrrrrl}
	\caption[]{Extracted WISE fluxes for the main-belt asteroid (10920).}\\
      \hline \noalign{\smallskip} \hline \noalign{\smallskip}
	JD$^{a}$ & $\lambda$ [$\mu$m]$^{b}$ & flux [mJy]$^{c}$ & error [mJy]$^{d}$ & r [au]$^{e}$ & $\Delta$ [au]$^{f}$ & $\alpha$ [$^{\circ}$]$^{g}$ & Band$^{h}$ & Comments$^{i}$ \\
      \noalign{\smallskip} \hline \noalign{\smallskip} \endfirsthead
      \caption[]{\emph{continued}} \\
      \hline \noalign{\smallskip} \hline \noalign{\smallskip}
	JD$^{a}$ & $\lambda$ [$\mu$m]$^{b}$ & flux [mJy]$^{c}$ & error [mJy]$^{d}$ & r [au]$^{e}$ & $\Delta$ [au]$^{f}$ & $\alpha$ [$^{\circ}$]$^{g}$ & Band$^{h}$ & Comments$^{i}$ \\
      \noalign{\smallskip} \hline \noalign{\smallskip} \endhead
      \hline
      \multicolumn{9}{c}{\emph{continued on next page}}
      \endfoot
      \noalign{\smallskip} \hline \noalign{\smallskip} \endlastfoot
      \label{app:wise_fluxes}
2455211.99146 & 11.10 &  56.043 &   4.070 & 3.0453 & 2.8596 &  18.8 & W3 & A,QFr10  	 \\ 
2455212.12390 & 11.10 &  53.471 &   3.870 & 3.0455 & 2.8579 &  18.8 & W3 & A,QFr10  	 \\ 
2455212.25620 & 11.10 &  26.947 &   2.010 & 3.0457 & 2.8562 &  18.8 & W3 & A,QFr10  	 \\ 
2455212.38851 & 11.10 &  50.877 &   3.682 & 3.0459 & 2.8545 &  18.8 & W3 & A,QFr10  	 \\ 
2455212.45472 & 11.10 &  29.959 &   2.226 & 3.0460 & 2.8536 &  18.8 & W3 & A,QFr10  	 \\ 
2455212.52081 & 11.10 &  59.940 &   4.354 & 3.0461 & 2.8528 &  18.8 & W3 & A,QFr10  	 \\ 
2455212.52094 & 11.10 &  58.307 &   4.207 & 3.0461 & 2.8528 &  18.8 & W3 & A,QFr10  	 \\ 
2455212.58703 & 11.10 &  46.915 &   3.407 & 3.0462 & 2.8519 &  18.8 & W3 & A,QFr10  	 \\ 
2455212.71933 & 11.10 &  62.190 &   4.473 & 3.0464 & 2.8502 &  18.8 & W3 & A,QFr10  	 \\ 
2455212.85176 & 11.10 &  36.385 &   2.652 & 3.0466 & 2.8485 &  18.8 & W3 & A,SAASEP=-3 QFr10 \\ 
2455212.98407 & 11.10 &  39.493 &   2.878 & 3.0468 & 2.8468 &  18.8 & W3 & A,QFr10  	 \\ 
\noalign{\smallskip}									      
2455379.80106 & 11.10 &  32.071 &   2.373 & 3.2857 & 3.0205 & -17.9 & W3 & A,QFr10  	 \\ 
2455379.93336 & 11.10 &  39.095 &   2.859 & 3.2859 & 3.0225 & -17.9 & W3 & A,QFr10  	 \\ 
2455380.06566 & 11.10 &  23.644 &   1.779 & 3.2860 & 3.0245 & -17.9 & W3 & A,QFr10  	 \\ 
2455380.19797 & 11.10 &  26.749 &   2.013 & 3.2862 & 3.0266 & -17.9 & W3 & A,QFr10  	 \\ 
2455380.26406 & 11.10 &  23.973 &   1.796 & 3.2863 & 3.0276 & -17.9 & W3 & A,SAASEP=+10 QFr10\\ 
2455380.33014 & 11.10 &  39.932 &   2.910 & 3.2864 & 3.0286 & -17.9 & W3 & A,SAASEP=+2 QFr10 \\ 
2455380.33027 & 11.10 &  39.059 &   2.878 & 3.2864 & 3.0286 & -17.9 & W3 & A,SAASEP=+3 QFr10 \\ 
2455380.46245 & 11.10 &  22.936 &   1.726 & 3.2866 & 3.0306 & -17.9 & W3 & A,QFr10  	 \\ 
2455380.46257 & 11.10 &  25.057 &   1.861 & 3.2866 & 3.0306 & -17.9 & W3 & A,QFr10  	 \\ 
2455380.52866 & 11.10 &  37.646 &   2.754 & 3.2867 & 3.0316 & -17.9 & W3 & A,QFr10  	 \\ 
2455380.59475 & 11.10 &  20.518 &   1.573 & 3.2868 & 3.0326 & -17.9 & W3 & A,QFr10  	 \\ 
2455380.66097 & 11.10 &  25.782 &   1.915 & 3.2868 & 3.0336 & -17.9 & W3 & A,QFr10  	 \\ 
2455380.79327 & 11.10 &  19.112 &   1.459 & 3.2870 & 3.0357 & -17.9 & W3 & A,QFr10  	 \\ 
2455380.92557 & 11.10 &  35.952 &   2.660 & 3.2872 & 3.0377 & -17.9 & W3 & A,QFr10  	 \\ 
2455381.05788 & 11.10 &  30.153 &   2.259 & 3.2874 & 3.0397 & -17.9 & W3 & A,QFr10  	 \\ 
\noalign{\smallskip}
2455211.99146 & 22.64 & 163.410 &  12.593 & 3.0453 & 2.8596 &  18.8 & W4 & A,QFr10  	 \\ 
2455212.12390 & 22.64 & 164.770 &  12.698 & 3.0455 & 2.8579 &  18.8 & W4 & A,QFr10  	 \\ 
2455212.25620 & 22.64 &  92.572 &   7.245 & 3.0457 & 2.8562 &  18.8 & W4 & A,QFr10  	 \\ 
2455212.38851 & 22.64 & 149.307 &  11.450 & 3.0459 & 2.8545 &  18.8 & W4 & A,QFr10  	 \\ 
2455212.45472 & 22.64 &  96.579 &   8.613 & 3.0460 & 2.8536 &  18.8 & W4 & A,QFr10  	 \\ 
2455212.52081 & 22.64 & 166.755 &  12.494 & 3.0461 & 2.8528 &  18.8 & W4 & A,QFr10  	 \\ 
2455212.52094 & 22.64 & 184.536 &  14.759 & 3.0461 & 2.8528 &  18.8 & W4 & A,QFr10  	 \\ 
2455212.58703 & 22.64 & 140.243 &  11.216 & 3.0462 & 2.8519 &  18.8 & W4 & A,QFr10  	 \\ 
2455212.71933 & 22.64 & 178.024 &  13.789 & 3.0464 & 2.8502 &  18.8 & W4 & A,QFr10  	 \\ 
2455212.85176 & 22.64 & 114.942 &   9.809 & 3.0466 & 2.8485 &  18.8 & W4 & A,SAASEP=-3 QFr10 \\ 
2455212.98407 & 22.64 & 122.484 &  10.022 & 3.0468 & 2.8468 &  18.8 & W4 & A,QFr10  	 \\ 
\noalign{\smallskip}
2455379.80106 & 22.64 & 122.710 &   9.869 & 3.2857 & 3.0205 & -17.9 & W4 & A,QFr10  	 \\ 
2455379.93336 & 22.64 & 123.617 &   9.434 & 3.2859 & 3.0225 & -17.9 & W4 & A,QFr10  	 \\ 
2455380.06566 & 22.64 &  82.886 &   6.782 & 3.2860 & 3.0245 & -17.9 & W4 & A,QFr10  	 \\ 
2455380.19797 & 22.64 &  96.135 &   8.359 & 3.2862 & 3.0266 & -17.9 & W4 & A,QFr10  	 \\ 
2455380.26406 & 22.64 &  78.719 &   7.204 & 3.2863 & 3.0276 & -17.9 & W4 & A,SAASEP=+10 QFr10\\ 
2455380.33014 & 22.64 & 122.710 &   9.760 & 3.2864 & 3.0286 & -17.9 & W4 & A,SAASEP=+2 QFr10 \\ 
2455380.33027 & 22.64 & 126.148 &  10.833 & 3.2864 & 3.0286 & -17.9 & W4 & A,SAASEP=+3 QFr10 \\ 
2455380.46245 & 22.64 &  80.627 &   7.010 & 3.2866 & 3.0306 & -17.9 & W4 & A,QFr10  	 \\ 
2455380.46257 & 22.64 &  75.942 &   6.950 & 3.2866 & 3.0306 & -17.9 & W4 & A,QFr10  	 \\ 
2455380.52866 & 22.64 & 131.486 &   9.987 & 3.2867 & 3.0316 & -17.9 & W4 & A,QFr10  	 \\ 
2455380.59475 & 22.64 &  75.107 &   6.370 & 3.2868 & 3.0326 & -17.9 & W4 & A,QFr10  	 \\ 
2455380.66097 & 22.64 &  79.374 &   7.554 & 3.2868 & 3.0336 & -17.9 & W4 & A,QFr10  	 \\ 
2455380.79327 & 22.64 &  67.746 &   6.321 & 3.2870 & 3.0357 & -17.9 & W4 & A,QFr10  	 \\ 
2455380.92557 & 22.64 & 128.493 &  10.513 & 3.2872 & 3.0377 & -17.9 & W4 & A,QFr10  	 \\ 
2455381.05788 & 22.64 & 102.443 &   8.382 & 3.2874 & 3.0397 & -18.0 & W4 & A,QFr10  	 \\ 
\noalign{\smallskip}
2455211.99146 &  4.60 &   0.299 &   0.072 & 3.0453 & 2.8596 &  18.8 & W2 & ISU,B,QFr10  	     \\ 
2455212.45472 &  4.60 &   0.139 &   0.050 & 3.0460 & 2.8536 &  18.8 & W2 & ISU,B,QFr10	     \\ 
2455212.52081 &  4.60 &   0.146 &   0.052 & 3.0461 & 2.8528 &  18.8 & W2 & ISU,B,QFr10	     \\ 
2455212.52094 &  4.60 &   0.297 &   0.068 & 3.0461 & 2.8528 &  18.8 & W2 & ISU,B,QFr10  	     \\ 
2455212.58703 &  4.60 &   0.171 &   0.053 & 3.0462 & 2.8519 &  18.8 & W2 & ISU,B,QFr10	     \\ 
2455212.71933 &  4.60 &   0.167 &   0.049 & 3.0464 & 2.8502 &  18.8 & W2 & ISU,B,QFr10	     \\ 
2455379.93336 &  4.60 &   0.277 &   0.063 & 3.2859 & 3.0225 & -17.9 & W2 & ISU,B,QFr10  	     \\ 
2455380.33027 &  4.60 &   0.221 &   0.063 & 3.2864 & 3.0286 & -17.9 & W2 & ISU,B,SAASEP=+3 QFr10 \\ 
\noalign{\smallskip}
2456674.91650 &  4.60 &   0.257 &   0.084 & 2.8588 & 2.6455 & -20.1 & W2 & NIS,B,QFr5  	     \\ 
2456675.04830 &  4.60 &   0.198 &   0.063 & 2.8587 & 2.6472 & -20.1 & W2 & NIS,B,QFr10 	     \\ 
2456675.31176 &  4.60 &   0.214 &   0.059 & 2.8583 & 2.6505 & -20.1 & W2 & NIS,B,QFr10 	     \\ 
2456675.31189 &  4.60 &   0.217 &   0.058 & 2.8583 & 2.6505 & -20.1 & W2 & NIS,B,QFr10 	     \\ 
2456675.64118 &  4.60 &   0.219 &   0.061 & 2.8580 & 2.6547 & -20.1 & W2 & NIS,B,QFr10 	     \\ 
2456675.90477 &  4.60 &   0.215 &   0.077 & 2.8576 & 2.6580 & -20.1 & W2 & NIS,B,QFr10	     \\ 
2456676.03657 &  4.60 &   0.297 &   0.070 & 2.8575 & 2.6597 & -20.1 & W2 & NIS,B,QFr10	     \\ 
2456974.42707 &  4.60 &   0.438 &   0.102 & 2.7188 & 2.5230 &  21.4 & W2 & NIS,B,QFr5	     \\ 
2456974.55848 &  4.60 &   0.449 &   0.106 & 2.7188 & 2.5213 &  21.3 & W2 & NIS,B,QFr10	     \\ 
2456974.55861 &  4.60 &   0.331 &   0.093 & 2.7188 & 2.5213 &  21.3 & W2 & NIS,B,QFr10	     \\ 
2456974.82156 &  4.60 &   0.315 &   0.101 & 2.7189 & 2.5179 &  21.3 & W2 & NIS,B,QFr10	     \\ 
2456974.95297 &  4.60 &   0.718 &   0.194 & 2.7190 & 2.5162 &  21.3 & W2 & NIS,B,QFr10  	     \\ 
2456975.15022 &  4.60 &   1.671 &   0.361 & 2.7191 & 2.5136 &  21.3 & W2 & NIS,B,QFr10  	     \\ 
2456977.97713 &  4.60 &   0.521 &   0.094 & 2.7202 & 2.4770 &  21.3 & W2 & NIS,B,QFr5	     \\ 
2456978.37162 &  4.60 &   0.364 &   0.091 & 2.7203 & 2.4719 &  21.3 & W2 & NIS,B,QFr10	     \\ 
2456978.50304 &  4.60 &   0.325 &   0.071 & 2.7204 & 2.4702 &  21.3 & W2 & NIS,B,QFr10	     \\ 
2456978.56874 &  4.60 &   0.327 &   0.075 & 2.7204 & 2.4694 &  21.3 & W2 & NIS,B,QFr10	     \\ 
2456978.70028 &  4.60 &   0.369 &   0.082 & 2.7205 & 2.4677 &  21.3 & W2 & NIS,B,QFr10	     \\ 
2456978.76599 &  4.60 &   0.276 &   0.063 & 2.7205 & 2.4668 &  21.3 & W2 & NIS,B,QFr10	     \\ 
2456978.83170 &  4.60 &   0.193 &   0.059 & 2.7205 & 2.4660 &  21.3 & W2 & NIS,B,QFr10	     \\ 
2456979.09465 &  4.60 &   0.391 &   0.095 & 2.7206 & 2.4626 &  21.3 & W2 & NIS,B,QFr5	     \\ 
2456979.22619 &  4.60 &   0.341 &   0.071 & 2.7207 & 2.4609 &  21.2 & W2 & NIS,B,QFr10	     \\ 
\noalign{\smallskip}
2457138.42474 &  4.60 &   0.394 &   0.095 & 2.8500 & 2.4988 & -20.4 & W2 & NIS,B,QFr10	     \\ 
2457141.57509 &  4.60 &   0.258 &   0.069 & 2.8537 & 2.5436 & -20.5 & W2 & NIS,B,QFr10	     \\ 
2457141.57522 &  4.60 &   0.246 &   0.085 & 2.8537 & 2.5436 & -20.5 & W2 & NIS,B,QFr10	     \\ 
2457141.96895 &  4.60 &   0.243 &   0.073 & 2.8541 & 2.5492 & -20.5 & W2 & NIS,B,QFr5	     \\ 
2457142.23152 &  4.60 &   0.262 &   0.078 & 2.8544 & 2.5530 & -20.5 & W2 & NIS,B,QFr5	     \\ 
2457142.69095 &  4.60 &   0.189 &   0.061 & 2.8550 & 2.5596 & -20.5 & W2 & NIS,B,QFr10	     \\ 
\noalign{\smallskip}
2458725.70223 &  4.60 &   0.217 &   0.061 & 2.8861 & 2.6835 &  20.5 & W2 & NIS,B,QFr10	     \\ 
2458725.96379 &  4.60 &   0.233 &   0.064 & 2.8858 & 2.6797 &  20.5 & W2 & NIS,B,QFr10	     \\ 
2458726.09469 &  4.60 &   0.224 &   0.076 & 2.8856 & 2.6778 &  20.5 & W2 & NIS,B,QFr10	     \\ 
2458726.16014 &  4.60 &   0.295 &   0.085 & 2.8855 & 2.6768 &  20.5 & W2 & NIS,B,QFr10	     \\ 
2458726.22547 &  4.60 &   0.288 &   0.079 & 2.8854 & 2.6759 &  20.5 & W2 & NIS,B,QFr5	     \\ 
2458726.42182 &  4.60 &   0.224 &   0.067 & 2.8852 & 2.6730 &  20.5 & W2 & NIS,B,QFr5	     \\ 
2458726.81428 &  4.60 &   0.226 &   0.075 & 2.8847 & 2.6674 &  20.5 & W2 & ISC,B,QFr5  	     \\ 
2458726.94505 &  4.60 &   0.230 &   0.074 & 2.8845 & 2.6655 &  20.5 & W2 & NIS,B,QFr5	     \\ 
\noalign{\smallskip}
2458875.40095 &  4.60 &   0.464 &   0.085 & 2.7412 & 2.2133 & -19.4 & W2 & NIS,B,QFr10	     \\ 
2458875.53185 &  4.60 &   0.315 &   0.073 & 2.7411 & 2.2149 & -19.4 & W2 & NIS,B,QFr10	     \\ 
2458875.66263 &  4.60 &   0.175 &   0.056 & 2.7411 & 2.2164 & -19.5 & W2 & NIS,B,QFr10  	     \\ 
2458875.79340 &  4.60 &   0.285 &   0.068 & 2.7410 & 2.2180 & -19.5 & W2 & NIS,B,QFr10	     \\ 
2458875.85885 &  4.60 &   0.213 &   0.067 & 2.7409 & 2.2187 & -19.5 & W2 & NIS,B,QFr10  	     \\ 
2458875.92418 &  4.60 &   0.504 &   0.102 & 2.7409 & 2.2195 & -19.5 & W2 & NIS,B,QFr10	     \\ 
2458875.92431 &  4.60 &   0.500 &   0.091 & 2.7409 & 2.2195 & -19.5 & W2 & NIS,B,QFr5	     \\ 
2458875.98912 &  4.60 &   0.230 &   0.076 & 2.7408 & 2.2203 & -19.5 & W2 & NIS,B,QFr10	     \\ 
2458876.12053 &  4.60 &   0.449 &   0.080 & 2.7408 & 2.2219 & -19.5 & W2 & NIS,A,QFr10	     \\ 
2458876.25131 &  4.60 &   0.353 &   0.072 & 2.7407 & 2.2234 & -19.5 & W2 & NIS,B,QFr10	     \\ 
2458876.38209 &  4.60 &   0.236 &   0.062 & 2.7406 & 2.2250 & -19.6 & W2 & ISC,B,QFr10  	     \\ 
2459199.33575 &  4.60 &   0.319 &   0.070 & 2.8360 & 2.6434 &  20.3 & W2 & NIS,B,QFr5	     \\ 
2459199.46653 &  4.60 &   0.190 &   0.063 & 2.8362 & 2.6417 &  20.3 & W2 & NIS,B,QFr5            \\ 
2459199.85886 &  4.60 &   0.311 &   0.079 & 2.8366 & 2.6367 &  20.3 & W2 & NIS,B,SAASEP=+0 QFr5  \\ 
2459199.92418 &  4.60 &   0.220 &   0.065 & 2.8367 & 2.6359 &  20.3 & W2 & NIS,B,SAASEP=-6 QFr10 \\ 
2459200.05445 &  4.60 &   0.265 &   0.076 & 2.8368 & 2.6343 &  20.3 & W2 & NIS,B,QFr10	     \\ 
2459200.31651 &  4.60 &   0.253 &   0.069 & 2.8371 & 2.6309 &  20.3 & W2 & NIS,B,QFr10	     \\ 
2459200.44728 &  4.60 &   0.241 &   0.065 & 2.8373 & 2.6293 &  20.3 & W2 & NIS,B,QFr10	     \\ 
2459200.57806 &  4.60 &   0.255 &   0.063 & 2.8374 & 2.6276 &  20.3 & W2 & NIS,B,QFr5	     \\ 
\noalign{\smallskip}
2459348.02845 &  4.60 &   0.172 &   0.049 & 3.0301 & 2.4990 & -18.0 & W2 & NIS,B,QFr5	     \\ 
      \noalign{\smallskip}
      \hline  
\end{longtable}  
  \footnotesize{
        $^{a}$ The observation epoch (Julian date);
        $^{b}$ the reference wavelength in the given bands (in $\mu$m);
        $^{c}$ the color-corrected, monochromatic flux densities at the reference wavelength
               (in mJy);
        $^{d}$ the absolute flux errors (in mJy);
        $^{e}$ the helio-centric distance r (in au);
        $^{f}$ the observer-object distance (au);
        $^{g}$ the phase angle (in $^{\circ}$;
        $^{h}$ the band name;
        $^{i}$ quality comments: all measurements were not saturated and had photometry quality
               flags 'A' (SNR$\ge$10) or 'B' (3$\le$SNR$<$10). Cases with South-Atlantic-Anomaly (SAA) separations $\le$10$^{\circ}$ are
               flagged. Only quality frame (QFr) scores 5 or 10 were accepted. For W2 detections we also listed the WISE catalog comments
               "NIS" (NoInertialSource), "ISU" (InertialSourceUndecided) or "ISC" (InertialSourceContamination).}\\

With the current spin-shape solution it is unfortunately not possible to combine all
data over 12 year as the rotation period is not known with sufficient quality. However,
the high-SNR WISE W3 and W4 data from January and July 2010 can be phased very well.
They show a strong rotational variation (see Fig.~\ref{fig:obsmod_wise}), perfectly
consistent with the lightcurve-derived rotation period. These flux changes are
explained by an ellipsoidal shape model with a/b $\approx$ 1.5.
The dual-band data are nicely balanced before and after opposition. The radiometric
study resulted in a thermal inertia of 15\,J\,m$^{-2}$s$^{-0.5}$K$^{-1}$ which allows to unify the
observation-to-model ratio for both bands and all before/after opposition data from 2010
(reduced $\chi^2$ close to unity). The best-fit sizes are 14.7\,km for the spherical shape,
and 15.3\,km for the ellipsoidal model (size of an equal-volume sphere).
The ellipsoidal spin-shape solution was also used for the JWST observations, with a
rotational phasing directly connected to our 2022 lightcurve measurements
(see Fig.~\ref{fig:mba_lc}).

\begin{figure}[h!tb]
	\resizebox{9cm}{!}{\includegraphics{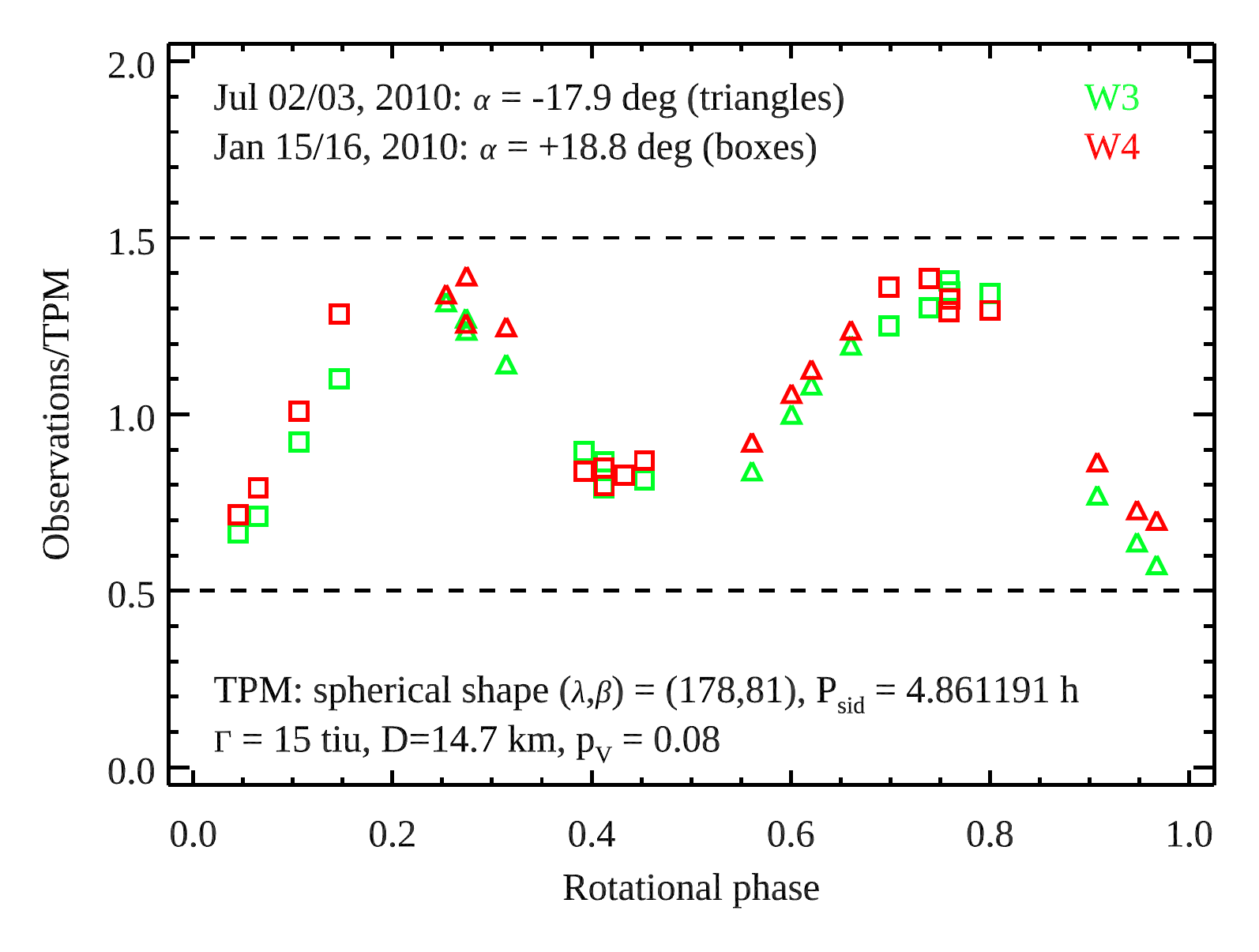}}
        \resizebox{9cm}{!}{\includegraphics{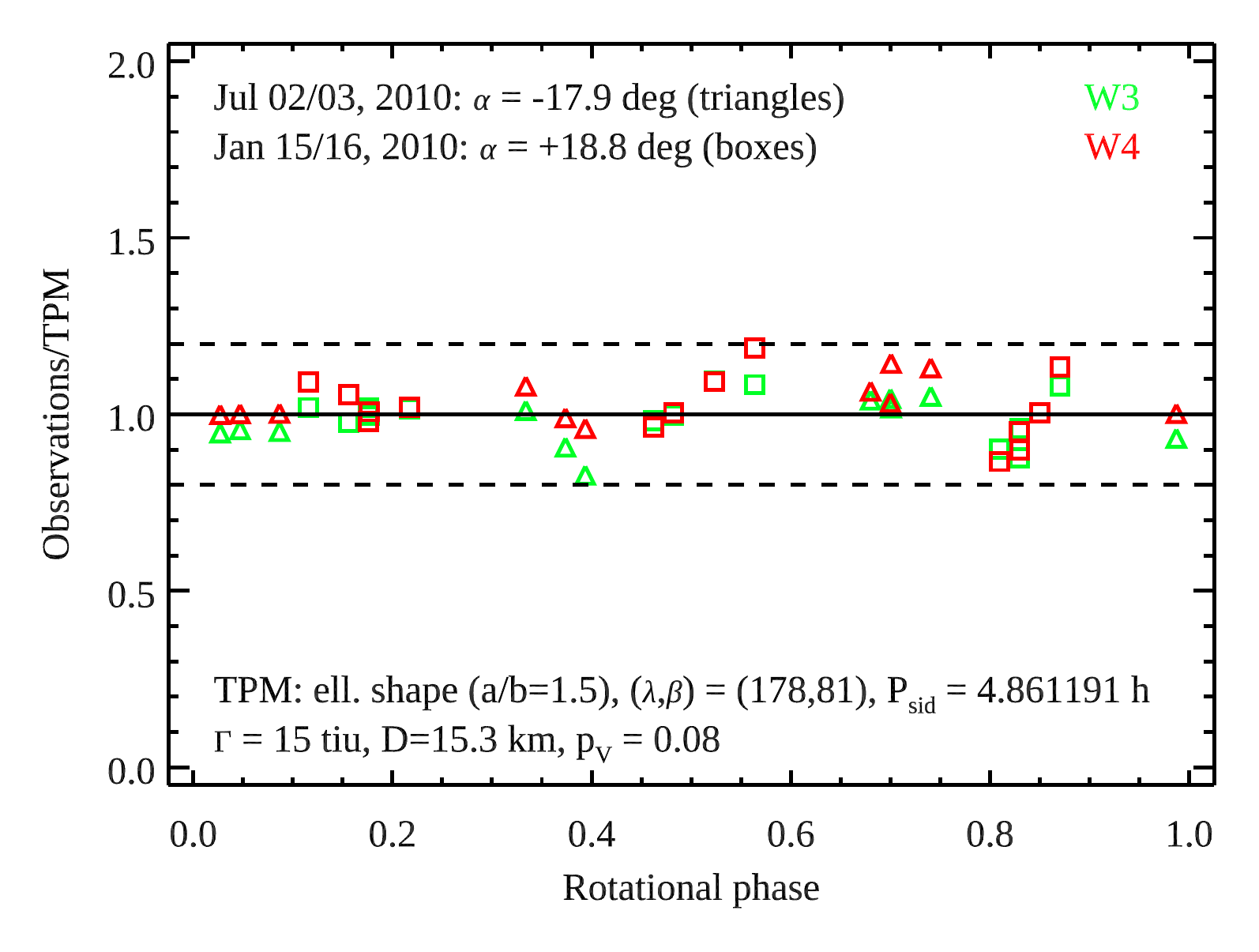}}
        \caption{WISE W3 and W4 data from 2010 from both epochs before and after opposition.
	         Left: WISE observations divided by a TPM prediction assuming a spherical
		 shape. Right: divided by a TPM prediction assuming an ellipsoidal shape
		 with a/b = 1.5. For both figures we take the same spin and thermal
		 properties into account.
         \label{fig:obsmod_wise}}
\end{figure}

\end{onecolumn}
\end{appendix}
	      
\end{document}